\documentclass[11pt,a4paper]{article}
\pdfoutput=1
\usepackage{jheppub}
\usepackage{graphicx}
\usepackage{natbib}
\usepackage{amsmath,amssymb}

\def\be{\begin{equation}}
\def\ee{\end{equation}}
\def\barr{\begin{array}{lr}}
\def\earr{\end{array}}
\def\bea{\begin{eqnarray}}
\def\eea{\end{eqnarray}}

\def\nn{\nonumber}

\def\ni{\noindent}

\def\del{\partial}
\def\delp{\del_+}
\def\delm{\del_-}

\def\sl2{$SL_2({\mathbb R})$}
\def\ha{\frac{1}{2}}
\def\sp{\sigma^+}
\def\sm{\sigma^-}

\def\a{\alpha}
\def\b{\beta}

\def\d{\delta}
\def\e{\epsilon}

\def\f{\phi}

\def\l{\lambda}

\def\q{\theta}
\def\r{\rho}
\def\s{\sigma}
\def\t{\tau}

\def\F{\Phi}

\def\L{\Lambda}

\title{\centering Closed Strings in the 2D Lorentzian Black Hole}
\author{\centering K. P. Yogendran}

\affiliation{\centering {\rm
  IISER Tirupati,\\
  Karakambadi Road, Tirupati,\\ Andhra Pradesh, 517507, India\\}
  {\it on deputation from}\\
  
  {\rm IISER Mohali, \\Sector 81 Knowledge City,\\
  Mohali, Punjab, 140306, India\\  }}

\emailAdd{pattag@gmail.com}

\abstract{ We revisit the spectrum of closed strings in the Lorentzian
  signature 2D black hole in string theory. Using the description of
  the black hole as a gauged WZNW model, we argue that the spectrum of
  the closed strings contain states from the spectrally flowed
  versions of the principal continuous and also the principal discrete
  series of \sl2. We identify the string configurations that
  correspond to these states. Using vector-axial duality, we also find
  new localized states that are essentially stringy in origin. }

\arxivnumber{1808.10109}
\begin{document}

\maketitle

\section{Introduction}
Recent years have seen great progress in understanding entropy and
grey-body factors of black holes using string theory. In one sense,
the problem of accounting for the microscopic states responsible for
the entropy has been solved. Not only the leading order
Bekenstein-Hawking entropy, but also an infinite series of subleading
corrections have been computed both in the microscopic description and
in the gravitational description and shown to match term by term (for
a review, see \cite{Sen-review}).

However, it is perhaps fair to say that a direct relationship between
the microstates that contribute to the entropy and the horizon, which
is the classical and geometrical manifestation of this entropy, has
not been demonstrated. This is, at least in part, due to the fact that
the counting of the degrees of freedom is performed at $G_N=0 $ when
the horizon vanishes.

The two dimensionsional black hole is particularly relevant in this
context. Not only is it an exact solution of string theory, but it
also admits a tractable CFT description in the form of a gauged WZNW
model.  Further, it has nonzero temperature and entropy
\cite{Nappi, Mann, Myers, Perry}.  Because of this one could hope to tie up the
relation between thermodynamics and the horizon more
explicitly. Another closely related reason for interest is that the
corresponding Euclidean black hole also admits an exact CFT
description. This fact should help in understanding the relationship
between the Euclidean and Lorentzian black holes.

Using the gauged sigma model description, we can study the spectrum of
the coset theory that corresponds to the Lorentzian black hole (in this
work, we concentrate on the Lorentzian sigma model). 
This was taken up in the paper of Dijkgraaf et. al
\cite{DVV}, and then a more detailed analysis appeared in the paper of
Distler and Nelson \cite{DistNel}. In the latter, the authors studied
the cohomology of the coset model in order to determine all {\em candidate}
physical states of the coset theory (from the hermitean
representations of the Kac-Moody algebra known at that time).

In spite of these investigations, the Minkowski black hole has not yet
been completely understood. In particular, we do not know how to
arrive at a modular invariant partition function (if that notion is
still relevant.
This is because of complications coming from the non-compact coset CFT
since the WZNW model is based on the group \sl2. In contrast, for the
Euclidean black hole the partition function is known \cite{Part}.

After the work of Maldacena and Ooguri \cite{mog}, it was
understood that string theory on $AdS_3$ (i.e., \sl2) requires
additional representations to form a modular invariant partition
function. These new representations maybe generated from the usual
ones by a transformation termed as ``spectral flow". Since the black
hole is obtained as a coset of the \sl2 CFT, a logical question is to
ask what happens to these new representations? Do they give rise to
new states of the black hole background?

Recall that the {\em Euclidean} black hole geometry looks like
$R\times S^1$ asymptotically. Therefore, we can have strings winding
on this circle. This winding number will however not be conserved
(since a winding string can "slip" off the tip). A short calculation,
presented in the appendix, shows that these winding strings appear as
the projection of the spectrally flowed representations of \sl2 to the
coset theory that describes the Euclidean black hole \cite{mog}. This
suggests that one should consider the spectrally flowed
representations in the Lorentzian case as well.

The Lorentzian black hole however, has no periodic direction and hence
no winding strings. Therefore, it appears that upon ``Wick rotation",
an entire tower of states disappears from the string spectrum!

Another motivation for searching for such states is the
following. There are D-braes in the Euclidean black hole geometry
which couple to the winding strings \cite{Ribault}. In \cite{KPY}, it
was shown that there are corresponding D-branes in the Lorentzian
black hole. In order to write the boundary states for these branes, we
can expect that we require the Lorentzian counterparts of winding
strings. Since the D-branes are not perturbative objects, this
provides a ``nonperturbative'' justification for searching for the
analogues of the winding strings. 

From the work of \cite{KKK}, it is known that the winding strings of
the Euclidean theory correspond to the non-singlet sector of the
matrix model which is the field theory dual (in the sense of
holography) to the black hole.  In \cite{Mal}, Maldacena studied
another set of nonsinglet modes in the form of folded ``long strings"
(which are dual to qq-states) in the asymptotically flat region of the
Lorentzian black hole. Therefore, we can ask how those strings lift to
the full black hole geometry (if at all).

However, to obtain the Lorentzian blackhole, we need to gauge a
hyperbolic direction in \sl2. Therefore we will need to understand the
action of spectral flow along this hyperbolic direction (in contrast
to \cite{mog}). This problem has already been studied by
Keski-Vakkuri and Hemming \cite{keski} in the context of BTZ black
holes. The BTZ black hole is an orbifold of \sl2, where the
orbifolding action is along a hyperbolic direction of \sl2 (i.e., we
orbifold by a boost). Thus, the string spectrum of the black hole will
contain states from the twisted sectors of the orbifold action. It was
shown by them that the twisted sector states may be understood as a
projection of spectrally flowed strings of the \sl2 theory, where the
spectral flow action is now along the hyperbolic direction. See also
\cite{Sugawara, Bars} for closely related explorations. 

Yet another perspective on the string theory of the black hole is
based on holography.  In the spirit of the AdS/CFT correspondence (and
further extensions along the lines of Vasiliev theory and the O(N)
model), we may regard the string theory of the black hole as a dual
description of the high temperature (deconfined) phase of a matrix
model that lives on the boundary. In this case the states of the
string theory correspond to states of the matrix model, while
operators of the matrix model should be dual to non-normalizable modes
of the bulk fields.


In our work, we will investigate the spectrum of the Lorentzian black
hole with careful attention to spectral flow. Rather than starting
with the spectral flow operation itself, we first consider geodesics
as representing point-like closed strings, and investigate if they
satisfy the physical state conditions of the string theory. By this
procedure, we are naturally led to spectrally flowed strings.  Thus
from the viewpoint of the 2D-black hole, we need to {\em start} with
these ``spectrally flowed'' representations of the \sl2 CFT.  We then
verify the earlier results about the spectrum of the string theory
with some interesting qualifications.  We show that the `tachyon'
occurs in a one parameter family, analogous to the tachyon in the
Euclidean black hole (in that case, the parameter is the winding
number). Thus we conclude that this new parameter (which arises as a
spectral flow parameter in both cases) is the Lorentzian equivalent of
the winding number. Secondly, we propose that the massless particle
corresponds to a {\em spectrally flowed} version of the coset primary.
Further, using the vector-axial self duality of the black hole sigma
model, we find additional, essentially stringy states by dualising the
geodesics. 

This manuscript is organized as follows. In Section \ref{LBH}, we
briefly discuss the black hole and its thermodynamic properties to be
self-contained. We will then recall its construction as a gauged sigma
model and discuss co-ordinate charts on \sl2 which project to various
regions of the black hole geometry in Section \ref{charts}.  A careful
reconsideration of the various geodesics of the black hole geometry in
Section \ref{Geodesics} gives us a handle on identifying the various
classical string excitations. Section \ref{VAD} uses the vector-axial
duality of this sigma model to produce new, essentially string states
which are `T-dual' to the point-like states of the previous section.
In section \ref{spectr}, we consider the physical state conditions,
both classical and quantum to be satisfied by the modes constructed
above. We conclude in Section \ref{summary} with a summary and some
suggestive directions for future research. Two appendices present
conventions employed in this paper and a short illustration of the
manner in which spectrally flowed representations descend as winding
strings of the Cigar geometry.

Part of the results in this paper has been published earlier, albeit
in condensed form \cite{Yog}. While this manuscript was being readied,
the paper \cite{sunny} appeared which also suggests an interior
structure to these black holes. 

\section{The Lorentzian black hole \label{LBH}}

The 2-D black hole is described by the following line element and
dilaton profile 
\be
ds^2=-(1-\frac{M}{r}) dt^2+\frac{k dr^2}{4 r^2(1-\frac{M}{r})}, 
\qquad e^\F=\sqrt{\frac{M}{r}} \label{schw}
\ee
It follows that, at the horizon $r=M$, the dilaton value is $\F=0$ and
that the string coupling blows up at the singularity $r=0$. The dimensionful parameter
$k$ is proportional to the level of the \sl2 coset theory described below. 
One can pass to global Kruskal co-ordinates by the change of variable
$u v=-\frac{(r-M)}{M}$ and $\exp(t)=-\frac{v}{u}$. The metric then becomes 
\be
ds^2=\frac{ -k\, du\, dv}{(1-uv)}
\ee
with the curvature singularity located at $uv=1$ and the horizon
at $uv=0$.

This black hole is a solution of the equations of motion that follow from the action
\be S= \int \sqrt{g} e^{-2\F}(R+4(\nabla \F)^2 +\L ), \label{Action} \ee
where $\L=-8$ is a negative cosmological constant.  Far away from the
black hole, the physics is controlled by the cosmological constant -
hence this black hole solution is analogous to a black hole inside
AdS-space \cite{Witten2}.

The thermodynamics of this black hole has been studied in various
works \cite{Nappi}-\cite{Mann} using various techniques. The black
hole has a finite temperature which can be easily determined by the
examining the periodicity of Euclidean time.  The ADM mass is read off
from the metric and then one can appeal to the first law of
thermodynamics $E=T S$ to obtain an entropy. These are 
\be
E=M, \quad T=\frac{1}{2\pi} \sqrt{\frac{M}{k}},  \quad S=2\pi\sqrt{Mk}
\ee 
Note that $M$ is completely determined by the dilaton value at the horizon 
$e^{\Phi(r=M)}=1 $.

\section{The Lorentzian black hole as a gauged sigma model}

The Lorentzian black hole is obtained by gauging a non-compact axial
$U(1)$ symmetry of the $SL(2,R)$ WZW model \cite{Wittenbh}. We shall
briefly outline the procedure here, for details refer to the appendix.

The symmetry that is being gauged corresponds to a hyperbolic subgroup
of \sl2 which acts on $g
= \left(\begin{array}{cc} a & u\\-v&b \end{array} \right)
\in$ \sl2, as
$\delta g=\epsilon (\sigma_3\,g+g\,\sigma_3)$, i.e.,
\bea\label{gauge-trafo}
\delta a &=& 2\epsilon a, \,\,\,\, \delta u = 0, \\\nonumber
\delta b &=& -2\epsilon b, \,\,\,\, \delta v = 0.
\eea
To obtain a target space interpretation, we have to gauge fix and
integrate out the gauge fields. In the region $(1-uv)>0$, we have $ab>0$, and
hence a natural gauge fixing condition is $ \label{gf} a=b.$ Upon
integrating out the gauge fields (which appear quadratically), we
obtain the black hole sigma model
\be
L=-\frac{k}{4\pi}\int d^2x\sqrt{h}\, \frac{h^{ij}\,\del_i u\,\del_j v
}{(1-u\,v)}
\ee
In the region $(1-uv)<0$ however, a good gauge fixing condition is
$a=-b$ (because $ab<0$). When $uv=1$, we have either $a=0$ or $b=0$ or both,
hence we cannot gauge transform a generic field configuration to the
gauge slice (for either gauge choice).  Although the gauge fixing
condition is singular, the sigma model is itself non-singular
everywhere (it has been argued that the locus $uv=1$ which maps to the
singularity is, by itself, governed by a topological string theory
\cite{Eguchi}). Requiring conformal invariance generates a dilaton
\be
\Phi=\Phi_0-\ha\ln (1-uv),
\ee
where the parameter $\Phi_0$ is related to the mass $M$ of the black
hole \ref{schw}.

\begin{figure}[t]
  \begin{center}
   \includegraphics[height=5cm]{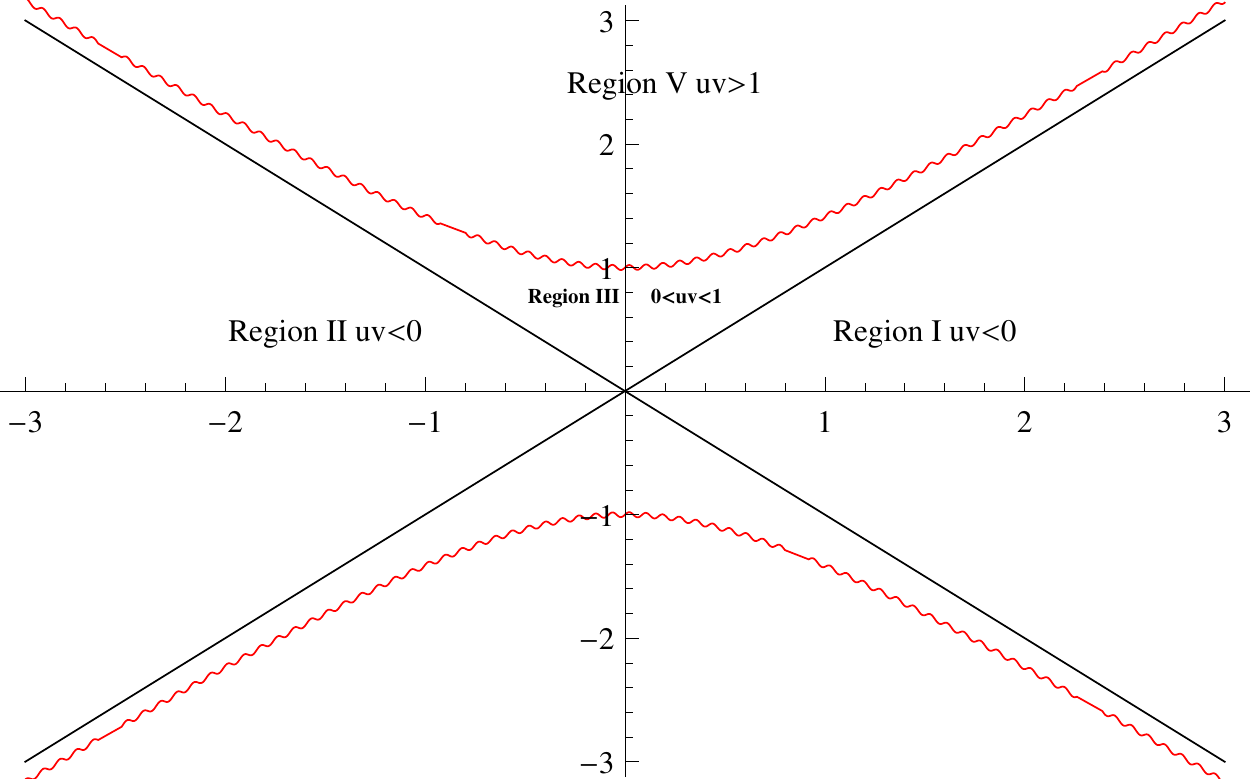}
  \end{center}
  \caption{The extended black hole geometry }\label{bhcover}
\end{figure}

The target space geometry of the sigma model so obtained is shown
in Fig: \ref{bhcover}.

In the figure, the diagonal lines $uv=0$ form the horizon, while
$uv=1$ is the singularity (the Ricci scalar diverges as $R\sim
(1-uv)^{-2}$). Regions I and II are asymptotically flat regions and in
regions V and VI time flows ``sideways''. 
Straight lines passing through the origin are constant time slices
with time increasing from top to bottom in region II, and from
bottom to top in region I. Thus the black hole singularity is in the
fourth quadrant (in the figure the diagonal lines are the $u,v$
co-ordinate axes!).

To obtain the physical state conditions and for quantization, it is 
convenient to follow the BRST procedure  \cite{DVV}. As a result of this, the 
black hole sigma model Lagrangian becomes 
\be
S_{BAH}=S'_{WZNW} (\r,t_L-\f_L,t_R-\f_R)- 
\frac{k}{2\pi}\int \delp X\delm X+S_{ghosts},
\label{Lbrst}
\ee
where $X=\f_L-\f_R$, and $A_\pm=\del_\pm\f_{L,R}$ are the gauge fields
of the gauged WZW model. The original gauge symmetry manifests itself as
invariance under a simultaneous shift of $t_{R,L}$ and $\f_{R,L}$. 
We note that $\r,t_{R,L}$ are the fields of the original WZNW model (before gauging) 
and hence are directly related to the target space variables of the black hole 
($t=\frac{1}{2}(t_R-t_L)$).
As a result of this procedure, we are left with a BRST constraint,
\be
k\del_\pm X= J'^{(2)} _\pm , \label{BRST}
\ee
where $J'$ in this equation is the conserved current of the WZNW Lagrangian S'
\ref{Lbrst} 
We also have a (classical) Virasoro constraint (since this is a string theory)
\be
T^{tot}_{++}=T_{++} ^{WZW'} -k(\del_+X)^2= 0 \label{Virasoro}
\ee
It is a remarkable fact that, instead of gauging the axial action as
above, if we gauge the vectorial action as in $\delta g=\epsilon
(\sigma_3\,g - g\,\sigma_3)$, we nevertheless obtain the same target
space geometry. In this case, the diagonal entries $a,b$ are invariant
under the gauging and form coordinates of the black hole spacetime.

\subsection{Co-ordinate systems for \sl2 \label{charts}}

It is useful to understand the correspondence between the
various regions of the black hole geometry and coordinate charts of \sl2.
First, we recall that every matrix $g \in$\sl2 with all entries nonzero can be
written as a product \cite{Vilenkin}
\be
g=d_1 (-e)^{\epsilon_1} s^{\epsilon_2} p \,d_2,
\ee
where $d_{1,2}={\rm diag}(e^{\theta_{1,2}}, e^{-\theta_{1,2}})$ and
$\,\,\,\theta_{1,2}\in(-\infty,\infty)$, $e$ is the identity matrix,
$s=i\s_2$ is the matrix $\left(\begin{array}{cc} 0&
1\\ -1&0\end{array}\right)$, $p$ is one of two matrices
\[ p_1=\left(\begin{array}{cc} \cosh\rho & -\sinh\rho\\
-\sinh\rho&\cosh\rho\end{array}\right),
\,\,\,\,\,\rho\in[-\infty,\infty),\quad
{\rm or}\quad
p_2=\left(\begin{array}{cc} \cos\rho &\sin\rho\\
-\sin\rho&\cos\rho\end{array} \right),
\,\,\,\,\,\rho\in[-\frac{\pi}{4},\frac{\pi}{4}], \]
and $\epsilon_{1,2}\in \{0,1\}$. Instead of using 4 trigonometric charts, we will extend the range
$-\frac{\pi}{2}\leq \r\leq \frac{\pi}{2}$ and drop the action of
$i\s_2$.
In a similar manner, matrices in \sl2 with at least one zero entry can be written as a product
\be
g=d\,(-e)^{\epsilon_1}\,s^{\epsilon_2}\,\left(\begin{array}{cc} 1 &
0\\ x & 1 \end{array} \right) s^{\epsilon_3}
\ee
where $d={\rm diag}(e^\phi,e^{-\phi})$.

The axial gauge symmetry that leads to the Lorentzian black hole acts as
$\theta_{1,2}\rightarrow \theta_{1,2}+\epsilon$, and the time
co-ordinate $t$ of the black hole geometry is related to the $\theta_i$
as $t=(\theta_1-\theta_2)$. It is then easy to see how the various
co-ordinate charts project down (upon gauging) to cover different
regions of the black hole geometry.

For instance, setting $p=p_1\,\epsilon_{1}=\e_2=0$ gives us \sl2
matrices of the form
\[g=d_1\left(\begin{array}{cc} \cosh\rho & \sinh\rho\\
  \sinh\rho&\cosh\rho\end{array}\right) \,d_2.\]
Gauging the axial $U(1)$ symmetry sets $d_2=d_1^{-1}=e^{-\s_3 t/2}$
and projects to the $(u,v)$ co-ordinates.  The matrices above are then
seen to cover the $uv<0$ regions of the black hole geometry with
$u=-e^{t}\sinh \r\quad v= e^{-t} \sinh\r$. We shall refer to the time
coordinate $t$ as `Schwarzschild' time and $\r\lessgtr0$ will correspond to
region I,II respectively.  Note that
$p=p_1,\,\epsilon_1=1,\,\epsilon_2=0$ also covers the same region of
the coset (for the full range of $\r \in \mathbb{R}$).

\begin{figure}
  \begin{center}
  \includegraphics[height=5cm]{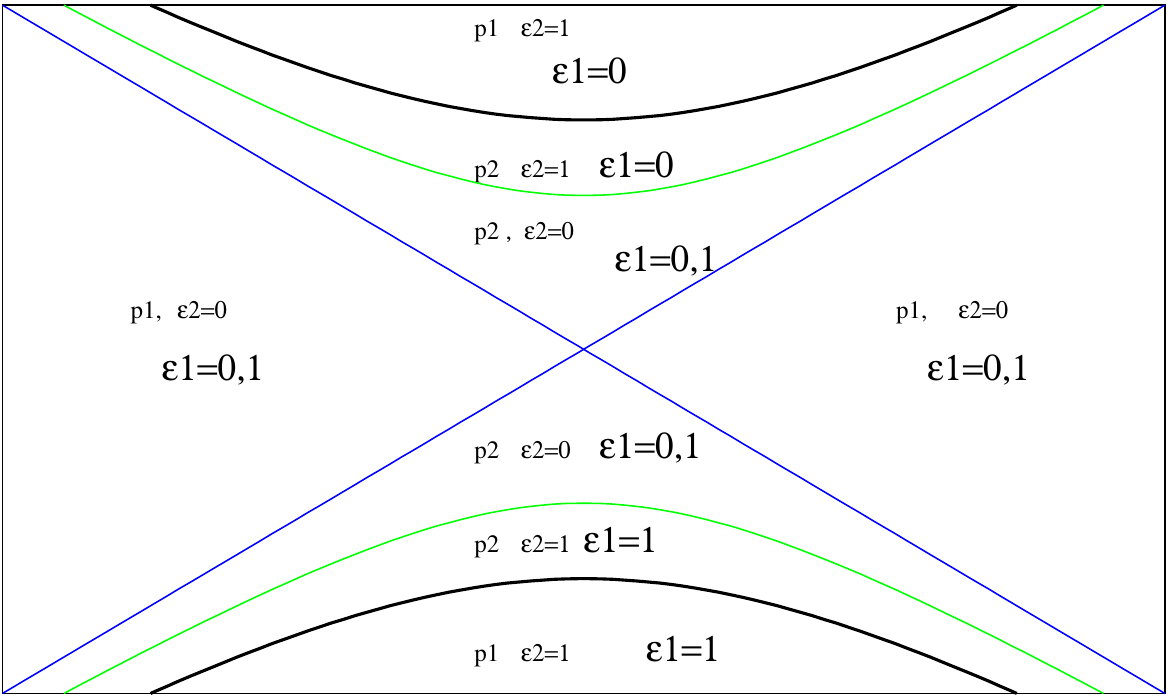}
  \caption{Covering diagram for the black hole geometry}
  \label{bh-newcover}
  \end{center}
\end{figure}

Thus we obtain the following covering diagram Fig. \ref{bh-newcover}.
The singularity ($ab=0$) is the dark (black) line in the figure, while
the horizon is the diagonal (blue) line $uv=0$. The matrices in \sl2
with zero entries cover the horizon lines and the singularity and are
not indicated in the figure.  The region between the horizon and the
singularity is covered by the four charts with $p_2$ type of matrices.

There are two $\mathbb{Z}_2$ operations which are important for our
purposes, multiplication by $-\mathbb{I}$ denoted as $R$ and $g \to
i\s_2\, g\, i\s_2$ denoted as $C$. $C$ reflects $u+v \to -(u+v)$ while
leaving $u-v$ unchanged - and $R$ is a simple reflection $(u,v) \to
(-u,-v)$.

In regions I and II, the $C$ operation is nothing but time reversal
(of the `Schwarzschild' time) and acts within each region. The $R$
operation maps region I to region II and vice versa.  In region V, the
chart is given by $u=e^{t}\cosh \r\quad v= e^{-t} \cosh\r$ and region
VI is obtained by the action of $R$.  However, in regions V and VI,
the $C$ transformation interchanges the regions besides reversing the
`Schwarzschild' time. In contrast to regions I and II, the full range
of $\r$ covers each region twice separately.

Regions III and IV are also covered twice by $u=e^{t}\sin \r, \quad v=
e^{-t} \sin\r$. $R$ and $C$ interchange regions III and IV, but $C$
reverses $t$ as well. The product $CR$, therefore, results in changing
the sign of $\r$ alone (which is the time direction in this region).

Thus, time reversal is implemented by $C$ in regions I and II, by $CR$
in regions III and IV, and by $R$ in regions V and VI. Also, to obtain
a single cover of the black hole, we can restrict to $\r \geq 0$, and
require $C$ and $R$ to be implemented faithfully on the states. It
must be noted that $C$ and $R$ are \sl2 transformations, while time
reversal is a property of the black hole geometry.

Thus for our choice of coordinate charts, straight lines passing
through the origin represent constant time surfaces in regions
I, II and V, VI.



\section{Geodesics and their lifts \label{Geodesics}}

The perturbative spectrum of the {\it Lorentzian} black hole has been
discussed in the literature (see \cite{DVV,DistNel}). The states in
the spectrum are a single massless scalar field, the ``tachyon'' in
the principal continuous series of \sl2 and possibly, some massive
states in the principal discrete series.  In the semiclassical limit,
the particles that correspond to these states will move on
null/timelike geodesics. Hence we may expect that the geodesics of
the geometry solve the equations of motion of the sigma model of the
black hole (the equations of the black hole sigma model reduce to
geodesic equations for point-like configurations - the dilaton also
plays a role only for stringy configurations).  Thus, one way to study
the spectrum would be to quantize the geodesics. It is of course
possible that some of the states of the string theory do not have a
classical limit i.e., are essentially ``stringy''.

However, rather than study the solutions to the sigma model defined by
the black hole metric, we will make use of gauged sigma model
description (see \cite{Bars} for a similar approach).  One advantage of
this description is that $\a'$ corrections can all be taken into account
(and for this black hole the $\frac{1}{k}$ corrections are quite large
\cite{Wittenbh}).  Because the black hole geometry is obtained by a
gauging procedure, the geodesics can be lifted to classical solutions
of the ungauged theory, viz. the \sl2 sigma model - upto gauge
ambiguities.

We shall discuss the various geodesics of the black hole geometry
(timelike geodesics and geodesic deviation for this black hole have
been discussed in \cite{sayan}). Uplifting these as solutions of the
\sl2 WZW model, we will find that the various geodesics all lift to
     {\em ``spectrally flowed''} geodesic solutions of \sl2. However,
     in this case, the various representations mix in a manner which
     is markedly different from that of the $AdS_3$ case \cite{mog}.

The geodesic equations (of the leading order geometry in $l_s$) are
\be
\dot \rho^2- \dot t^2\tanh^2\rho=\epsilon\qquad \dot t\tanh^2\rho=E
\ee
$\epsilon=0,\mp$ accordingly as we are considering null, time-like or
space-like geodesics. This form of the equations are relevant to the
asymptotically flat regions I and II of the preceding section.  We
shall discuss the other regions of the black hole geometry in in
order to determine multiplicities of the current algebra states in the
string theory spectrum.  For simplicity, we will focus on one example
in each case now (we will have occasion to examine the action of
symmetries on the geodesics later).

We will proceed as follows - we first determine the geodesic
solutions, and then construct the quantities $u=-e^{-t(\tau)}
\sinh\rho(\tau)$ and $v=e^{t(\tau)} \sinh\rho(\tau)$ (this is
appropriate for Region I of the geometry).

We can then form an \sl2 matrix as $\left(\barr a&u\\-v&b\earr\right)$
where $a,b=\sqrt{1-uv}$. The non-uniqueness of this procedure lies in
the fact that only the product $ab=1-uv$ is determined. This ambiguity
is of course a gauge artefact - imposing the gauge fixing condition
\ref{gf} will result in a unique lift.

We shall choose $a,b$ in such a manner that it is easy to factorize
the resultant \sl2 matrix into a product $g=g_+(\sp)\,g_-(\sm)$, and
thus manifestly a solution of the \sl2 CFT.  It may happen that there
are solutions of \sl2 theory which are gauge inequivalent, but which
nevertheless give rise to the same solution of the gauged model. In
that case, we must treat the various solutions separately.

\subsection{Null geodesics}
\begin{figure}
  \begin{center}
  \includegraphics[height=5cm]{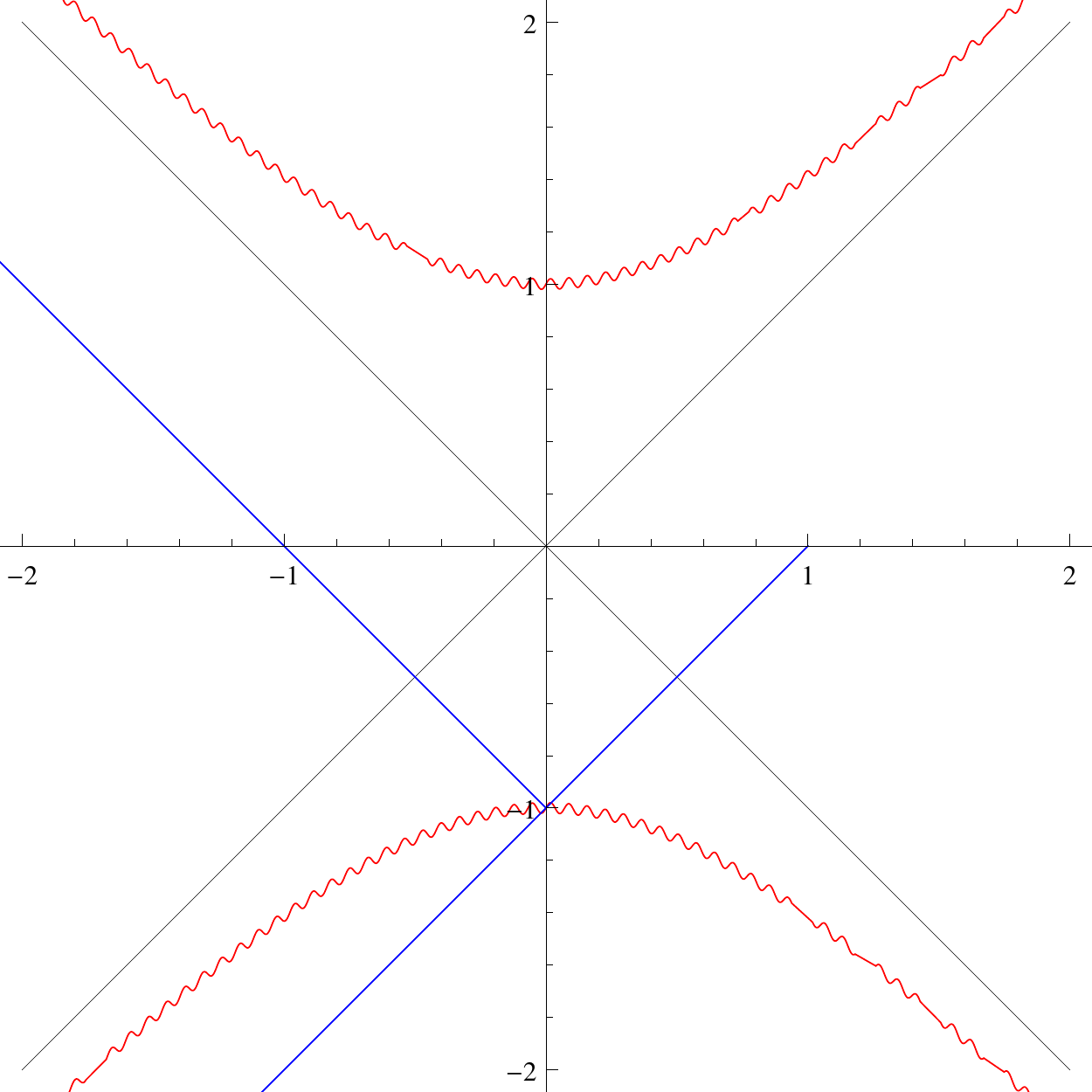}
         \caption[Example of figure]{Null geodesics.}%
         \label{null-geodesics}
  \end{center}
\end{figure}

The null geodesics of this geometry are the same as in flat two 
dimensional space (because the 2-D black hole is conformally 
flat). For instance, one family of null geodesic solutions are
\be u= -1, \quad v=e^{2E(\tau-\tau_0)}-1,  \ee
where $\tau$ is an affine parameter. For now, we set the integration
constant $\tau_0=0$, but the general cases will be discussed in
Section \ref{reps}. Some null geodesics are shown as blue straight
lines in Fig: \ref{null-geodesics}.

This solution can be written as an \sl2 matrix
\be
g=\left(\begin{array}{cc}
e^{E(\tau-\sigma)}&-1\\-(e^{2E\tau}-1)&e^{E(\tau+\sigma)}\earr\right),
\label{null-lift}
\ee
where, we have used the (gauge) ambiguity of the lifting procedure to introduce
sigma dependence in the $(1,1)$ and the $(2,2)$ entries. The above matrix is
not periodic in \sl2 - however, it does represent a {\it closed} string in
the {\em coset} theory because it is periodic upto a constant gauge transformation
(or more trivially, the {\em coset} solution has no $\sigma$-dependence).

This matrix $g$ can be factorized into a product of \sl2 matrices
\be
g=e^{-\frac{E\sp}{2}\sigma_3}\,g_+(\sp)g_-(\sm)
e^{\frac{E\sigma^-}{2}\sigma_3},
\ee
where
\[g_+(\sp)=
\left(\begin{array}{lr} e^{\frac{\b\sp}{2}} & 0 \\
-e^{\frac{\b\sp}{2}} & e^{-\frac{\b\sp}{2}}  \earr\right), 
\hspace{1cm} g_-(\sm)=
\left(\begin{array}{lr} e^{\frac{\b\sm}{2}} & -e^{\frac{\b\sm}{2}} \\
e^{-\frac{\b\sm}{2}} & 0 \earr\right).
\]
For the null geodesics under consideration, the parameter $\b$ should be set equal to $E$.

Since $g$ can be written as a product of a purely right-moving and a
purely left moving matrix, it solves the classical equations of motion
of the \sl2 WZNW model as well\footnote{A general solution to the
equations of motion of a WZNW model is a product of a matrix whose
entries are purely right moving with another matrix whose entries are
purely left-moving}.  The factorization above into {\it four} factors
might seem somewhat arbitrary since the product of the first two is
still purely right-moving and the product of the last two is purely
left-moving. However, note that the \sl2 matrix $\tilde g=g_+ g_-$ is
also a solution of the WZNW model and furthermore 
\be
\tilde g= \left(\barr 1&0\\-1&1\earr\right)
\left(\barr e^{\b\tau}& 0\\0& e^{-\b\tau}\earr\right)
\left(\barr 1&-1\\1&0\earr\right)
\label{gtn}
\ee
is a function of
$\tau$ alone (whereas the product of all four is not). 

Therefore the (null) geodesic is obtained in a two step procedure.
First consider the \sl2 solution $\tilde g$ above. This represents a
point-like particle trajectory in \sl2. Since the `proper time'
$(\frac{ds}{d\tau})^2>0$ (in \sl2) (our sign conventions are presented
in the appendix), $\tilde g$ represents a space-like {\em geodesic} in
\sl2.  We then construct a new matrix
\be g=\exp(\frac{w}{2}\s_3 \s_+)\, \tilde g\, \exp(-\frac{w}{2}\s_3 \s_-), \label{specflow}\ee
where $\s_\pm=\t\pm\s$ .  This operation leads to a new solution of the \sl2
conformal field theory.  This translation parallels the spectral flow
operation in \cite{mog} - the difference being that we perform the
flow along the noncompact direction of \sl2. However, this is the same
as the spectral flow relevant for the BTZ black hole \cite{keski}.

Since we are gauging the axial symmetry, the sigma dependence coming
from this spectral flow operation may be gauged away.  The off
diagonal entries of the resultant \sl2 matrix are invariant under the
gauging (\ref{gauge-trafo}) that leads to the black hole. Hence, they
can be projected to the black hole geometry and interpreted as string
configurations on the black hole.  In this manner, we obtain the full
solution after ``spectral flow''.  To obtain, the particular null
geodesic of the previous section, we must interpret $w$ above as $-E$.

Note that since $\tilde g$ is a pointlike trajectory in \sl2, it might
naturally be associated with the primaries of the \sl2 conformal
theory. This is because the vertex operators that correspond to
point-like solutions can be regarded as eigenfunctions of the
Laplacian operator on \sl2. The coset states are then obtained by a
projection condition.

The first thing to note is that the point-like \sl2 solution above
cannot be transformed by constant matrices into the null geodesic of
the black hole. As a consequence, the state corresponding to the null
geodesic of the black hole (``tachyon'') is not directly the primary
of the coset theory (associated with the spacelike geodesic in \sl2).

The second thing to note is that spectral flow in the hyperbolic
direction seems to produce a string in \sl2 that extends all along
x-direction (parallel to the boundary). Thus it would correspond to a
large excitation in the $AdS_3$ string theory. However, the current
algebra charges are all finite for this string. 


The third thing to note is that the pointlike geodesic in \sl2 was
written as the product $g_+ g_-$ where the $g_\pm$ were not periodic in
the $\sigma$-coordinate. This is not a serious shortcoming since the
pointlike trajectory of \sl2 maybe written using other forms for
$g_\pm$ - this will not alter the analysis, the key point being the
coset solution is being obtained {\em after} spectral flow and using
the {\em spacelike} geodesic in \sl2. 

For this solution, we can determine the left and right moving charges
$\tilde J^{(2)}_\pm$ of the current algebra, and also the world sheet
stress tensors $\tilde T_\pm$ (for conventions, see Appendix A). And
similar quantities can be calculated for the matrix $g$ (denoted
without a tilde).
These are as follows.
\bea
\tilde J^{(2)}_\pm = -\frac{k\b}{2} \,&;&
\,\, \tilde T_\pm= \frac{k}{4}\b^2, \\\nn
J^{(2)}_\pm =\frac{k}{2}(w-\b)&;&\,\,T_\pm=\frac{k}{4} (w-\b)^2, \label{T-null}
\eea
which obey the equations
\bea
J^{(2)}_\pm =\tilde J^{(2)}_\pm + k \frac{w}{2}, \\\nn
T_\pm=\tilde T_\pm+ w \tilde J^{(2)}_\pm + k\frac{w^2}{4},  \label{T-flow}
\eea
with $w$ being the spectral flow parameter. In contrast with the
spectral flow operation in \cite{mog}, since this operation is along
the $J^{(2)}$ direction, we have different signs in the above
equations.

We can also determine the $J^{(0)}$ quantum numbers for the above
solution (which is related to the energy in global AdS coordinates -
the $J^{(0)}$ direction is the compact elliptic direction in \sl2)
\be
\tilde J^{(0)}_\pm =\pm \frac{\b}{2} \qquad J^{(0)}_\pm =\pm \frac{\b}{2} e^{w\s_\pm}.
\ee
It is worth noting that before spectral flow, the matrix $\tilde g$ is
an `eigenstate' of $J^{(0)}$, while after spectral flow this is no
longer the case.


\subsection{The timelike geodesics}

From the geodesic equations, it immediately follows that we have three
families of massive geodesics depending on $E^2-m^2$. They also fall
into distinct representations of \sl2 when we uplift them to classical
solutions of the un-gauged theory. 

\ni{\bf Case 1: Geodesics with $E^2>m^2$}

These geodesic trajectories reach ${\mathcal J}^\pm$ at late (early) times.
They may thus be thought of as either particles falling into the black
hole or particles that scatter out to asymptotic infinity. An example is
\be
u=\frac{-e^{-E\tau}}{\sinh\f}\sinh(\beta\tau+\f)\qquad
v=\frac{e^{E\tau}}{\sinh\f}\sinh(\beta\tau-\f), \label{tl}
\ee
where $\beta=\sqrt{E^2-m^2}$, $\tanh^2\phi=\frac{\beta^2}{E^2}$.
These geodesics, shown as a pair of black dashed lines in Fig:
\ref{timelike-geodesics}, satisfy $u(-\t)=v(\t)$.
\begin{figure}
\begin{center}
  \includegraphics[height=6cm]{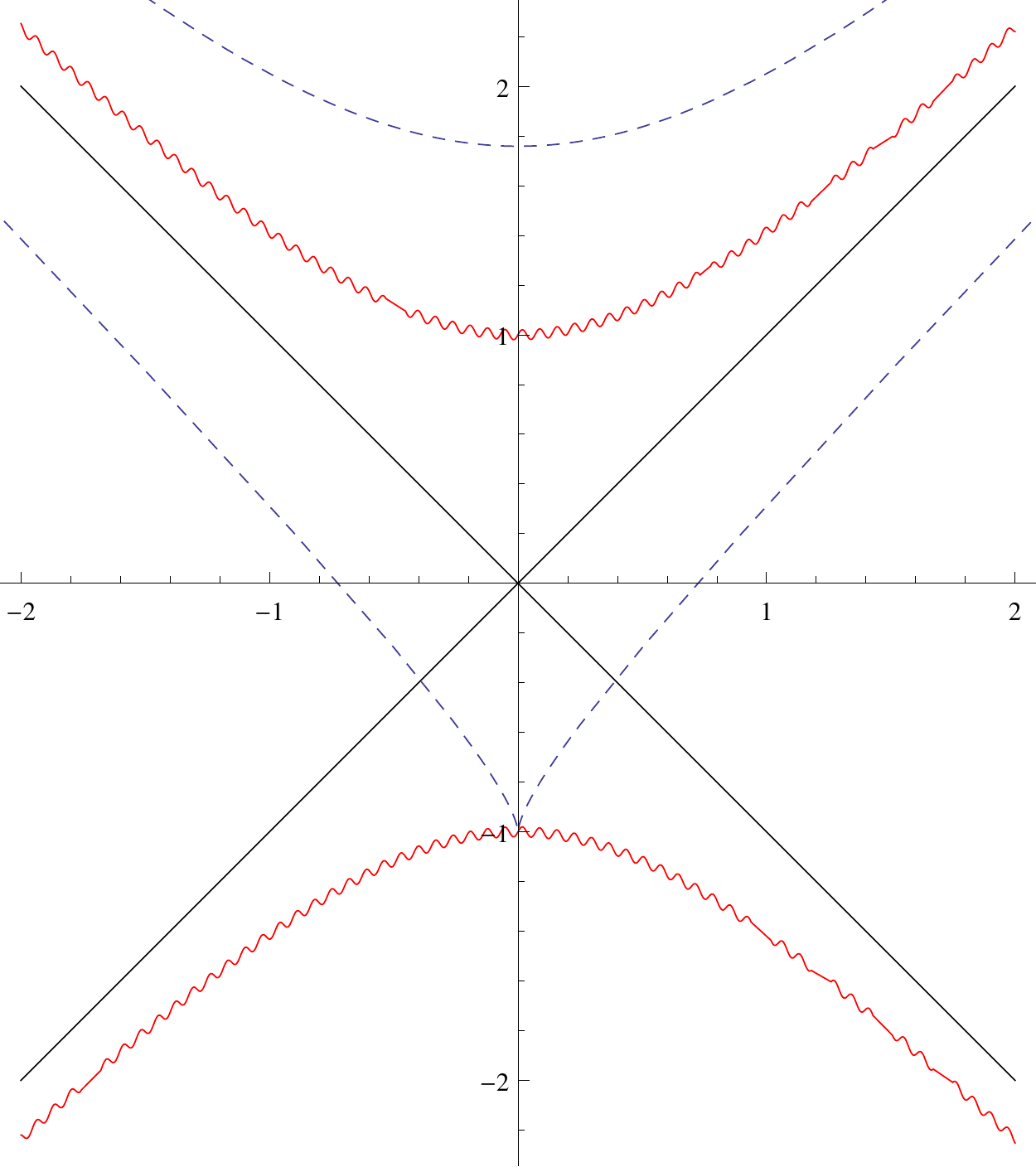}
         \caption[Example of figure]{Timelike geodesics.}
         \label{timelike-geodesics}
\end{center}
\end{figure}

This solution too can be lifted up to \sl2 by choosing matrices 
$\tilde g_\pm$
\[
\sqrt{\sinh3\phi\sinh\phi}\quad\tilde g_+(\sigma^+)=\begin{pmatrix}
\sinh(\ha \beta \sigma^+-\phi) & -\sinh(\ha \beta \sigma^+ +2\phi)\cr
-\sinh(\ha \beta \sigma^+-2\phi)&\sinh(\ha \beta \sigma^++\phi),\end{pmatrix}
\]
and
\[
\sqrt{\sinh3\phi\sinh\phi}\quad\tilde g_-(\sigma^-)=\begin{pmatrix}
\sinh(-\ha \beta\sigma^-+2\phi) & \sinh(\ha \beta\sigma^- -\phi)\cr
-\sinh(\ha \beta\sigma^-+\phi) & \sinh(\ha\beta\sigma^-+2\phi)
\end{pmatrix}.
\]
The full solution is obtained after spectral flow on $\tilde g$ as before \ref{specflow}. 
The product of the two matrices gives
\[\tilde g=
\frac{1}{\sinh\phi}\left(\barr \sinh\beta\tau&-\sinh(\beta\tau+\phi)\\
-\sinh(\beta\tau-\phi)& \sinh\beta\tau\earr\right).\]
It is easily shown that $\tilde g$ can be rewritten as $\tilde g=U
\left(\barr e^{\beta\tau}&0\\0&e^{-\beta\tau}\earr\right)V,$ where
$U,V$ maybe chosen to be
\[U=\frac{1}{\sqrt{2\sinh\phi}} \left(\barr 
e^{\frac{\f}{2}} & -e^{-\frac{\f}{2}}  \\
-e^{-\frac{\f}{2}}  & e^{\frac{\f}{2}}  \earr 
\right) {\rm and }\qquad
V=\frac{1}{\sqrt{2\sinh\phi}} \left(\barr 
e^{-\frac{\f}{2}}  & -e^{\frac{\f}{2}} \\
e^{\frac{\f}{2}}  & -e^{\frac{-\f}{2}} \earr\right).\]

Note that both the null and massive geodesics above are obtained from
the spacelike geodesics of \sl2 (before spectral flow).  Therefore,
one can transform the corresponding \sl2 matrices into each other by a
constant transformation in \sl2.


For this solution, we can calculate the Kac-Moody charges:
\bea
\tilde J^{(2)}_\pm = -\frac{k\,\b\coth\f}{2} \,&;&\qquad \tilde T_\pm= \frac{k}{4} \beta^2\\\nn
J^{(2)}_\pm =\frac{k}{2}(w-\b\coth\f)\,&;&\qquad T_\pm=\frac{k}{4}(\beta^2+w^2-2w\b\coth\f)
\label{timelike-I}
\eea
It may be observed that $\tilde T>0$ and $J^{(2)} _\pm> -k\b$ which
together may be interpreted as defining the principal continuous
representation of \sl2 (in our sign conventions) --- because $\tilde
T>0$ only for this representation and $J^{(2)}$ is bounded on one
side.

Using the above solution, we find 
\be
\cosh^2\r=\frac{\sinh^2 \b\t}{\sinh^2\f}\sim \frac{e^{2|\b|\t}}{\sinh^2\f}. \label{radial-momentum}
\ee
From the expressions above, we can see that $\b$, which is related to the quadratic
Casimir $T$ of \sl2 ,determines the momentum along the $\r$ direction. Also, 
\be
\exp(2t)=\exp(2w\t)\sqrt{\frac{\sinh(\b\t-\f)}{\sinh(\b\t+\f)}}\sim \exp(2w\t-2\f),
\ee
which shows, rather surprisingly, that the spectral flow parameter $w$
determines the energy of the state in the bulk. Note that if we had
interpreted the parameter $\b$ as the energy of the state, and if we
use the unitarity bound on the \sl2 current algebra representations
(of the continuous series), we would have concluded that the energy is
bounded above. This is now no longer an issue since $w$ is allowed to
be any real number. Henceforth, we will use $E$ to label the spectral
flow parameter instead of $w$. A related observation has already
appeared in the literature \cite{Israel}.

A last point for consideration is that the solution, Eq. \ref{tl}, for the
full range of the affine parameter $\t$, actually represents a pair
(see Fig: \ref{timelike-geodesics}) of geodesics, one each in regions
I and II.  If the one in region I (solid black curve in the fourth
quadrant) is interpreted as emanating from the singularity and going
to $\mathcal{J}^+$ (depending on the sign of $E$), then the other
(solid black curve in the second quadrant) in region II falls into the
singularity (and vice versa).  Both members of a pair intersect at the
singularity - which represents `the end of time'. There are, in fact,
no time-like geodesics near the singularity on the other side as we
will see later.


Under $C$ operation, which is time reversal, the solid curve in Fig:
\ref{timelike-geodesics} with parameters $(E,\b, J^2)$ is mapped to
the dashed curve with parameters $(-E,\b,-J^2)$.  The same thing
happens under the $R$ operation. 

In the \sl2 quantum theory, we therefore interpret the state
$|E,\b,J^2 \rangle$ as representing this pair of particles (and not a
single particle in isolation in region I).

Since both timelike and null geodesics are obtained from the same
parent \sl2 states by `spectral flow', an outgoing {\em null} geodesic
in region I must also be similarly `paired' with another ingoing null
geodesic with the same quantum numbers.

{\bf  Case 2: Geodesics with $E^2<m^2$}

When the energy is lower than the mass, the particle geodesics become
qualitatively different from the previous case. These solutions can be
interpreted as analytical continuations in $\b$, and therefore $\f$,
of the timelike geodesics of the previous section.
\be
u=-\frac{e^{-E\t}}{\sin\f}\sin(\beta\t+\f), \qquad v=\frac{e^{E\t}}{\sin\f}\sin(\beta\t-\f),
\ee
where $\beta^2+E^2=m^2$, and $\tan^2\f=\frac{\b^2}{E^2}$.  Since,
$1>uv>-\rm{cot}^2\f$, these geodesics never reach asymptotic infinity
and thus may be interpreted as bound states.


\begin{figure}
  \begin{center}
\includegraphics[height=5cm]{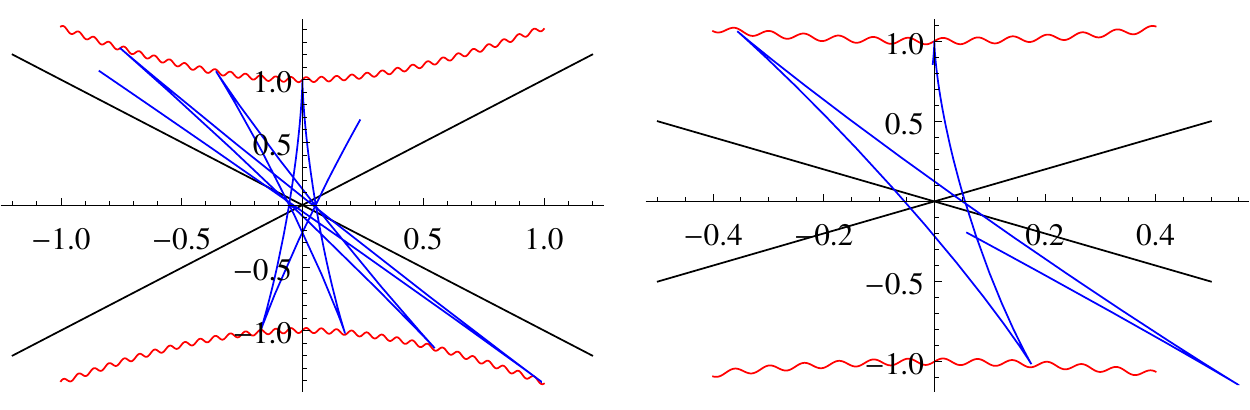}
         \caption[Example of figure]{A localized geodesic}
         \label{horiz-geodesics}
  \end{center}
  
\end{figure}

The \sl2 matrices in this case can also be obtained by analytic 
continuation
\[
\sqrt{\sin3\phi\sin\phi}\,\tilde g_+(\sigma^+)=\begin{pmatrix}
\sin(\ha \beta \sigma^+-\phi) & -\sin(\ha \beta \sigma^+ +2\phi)\cr
-\sin(\ha \beta \sigma^+-2\phi)&\sin(\ha \beta \sigma^++\phi)\end{pmatrix},
\]
and
\[
\sqrt{\sin3\phi\sin\phi}\,\tilde g_-(\sigma^-)=\begin{pmatrix}
\sin(-\ha \beta\sigma^-+2\phi) & \sin(\ha \beta\sigma^- -\phi)\cr
-\sin(\ha \beta\sigma^-+\phi) & \sin(\ha\beta\sigma^-+2\phi)\end{pmatrix},
\]
with $g=\exp(-\frac{E}{2}\s_3 \s_+)\, \tilde g\, \exp(\frac{E}{2}\s_3 \s_-)$

Again, the matrix $\tilde g=\tilde g_+ \tilde g_-$ can be written as a product
$U\, \begin{pmatrix}\cos\beta\tau &\sin\beta\tau\cr  -\sin\beta\tau&\cos\beta\tau
\end{pmatrix}\, V, $
with the following choices for $ U,V\in SL(2,R)$
$$
U=\frac{1}{\sqrt{\sin\phi}}\left(\barr 
\sin\phi/2 &-\cos\phi/2 \\ 
\sin\phi/2&\cos\phi/2 
\earr\right) \quad
V=\frac{1}{\sqrt{\sin\f}}\left(\barr
\cos\phi/2 &-\cos\phi/2 \\ 
\sin\phi/2&\sin\phi/2 
\earr\right).
$$
Hence, these are obtained by spectrally flowing {\em time-like}
geodesics of \sl2 as can be verified by computing the `proper time'
$\left(\frac{ds}{d\t}\right)^2 <0$ in \sl2.

For these solutions, we can calculate the Kac-Moody charges:
\bea
\tilde J^{(2)}_\pm = -\frac{k\b \cot\f}{2} \,&;&
\,\, \tilde T_\pm= -\frac{k}{4} \beta^2, \\\nn
J^{(2)}_\pm
=\frac{k}{2}(E-\b\cot\f)\,&;&\,\,
T_\pm=\frac{k}{4}(-\beta^2+E^2-2E\b\cot\f).
\eea
In this case, $\tilde T$ is bounded above, $\tilde T<0$, which means that these
geodesics fall into the discrete series representations of \sl2 (since
for these representations, the quadratic Casimir is bounded above).

Thus, we see that as we vary the energy of the massive string in the
black hole geometry, we pass between time-like and space-like
geodesics in \sl2.  However, all these map onto time-like geodesics of
the black hole geometry.  In \cite{mog}, it was observed that as one
increases the energy of strings in $AdS_3$, the parent
configurations passed from time-like geodesics to space-like geodesics
(i.e., short strings to long strings). The curious thing is that the
energy in global $AdS_3$ is related to the compact direction in \sl2 while
the energy of the black hole geometry is related to a hyperbolic
direction in \sl2. Yet, a similar thing happens in this case.

In \cite{mog}, it was argued that the spacelike geodesics of $AdS_3$ give
rise to the (spectrally flowed) principal continuous representations
of the \sl2 Kac-Moody algebra. These long strings were scattered out to
infinity.  In our case also, we observe that the strings obtained from
the spacelike geodesics of \sl2 can reach the future/past infinities -
and are thus visible in the asymptotic region of the black hole.
The discrete representations on the other hand gave rise to
states which are localized in the interior of  $AdS_3$. This is true
of the corresponding geodesics of the black hole geometry as well.


We can also construct geodesics with $E^2=m^2$
\bea
u&=&-e^{-\tau}(\tau+1)\\\nn
v&=&e^\tau(\tau-1)
\eea
In this case, $\cosh^2\r=\tau^2$, and 
\be
g=
\left(\barr e^{-\ha\sigma^+}& 0\\0 & e^{\ha\sigma^+}\earr\right)
\left(\barr \ha\sigma^+& 1 \\-\ha\sigma^+ +1 & -1\earr\right)
\left(\barr 1 & -1 \\\ha\sigma^- & -\ha\sigma^--1\earr\right)
\left(\barr e^{\ha\sigma^-}& 0\\0 & e^{-\ha\sigma^-}\earr\right)
\ee
These solutions above are unusual in the sense that the solution do
not seem to depend on any parameters at all.

\subsection{Spacelike geodesics}

In this case, the equations of motion have the following solution,
\be
u=-\frac{e^{-E\t}}{\cosh\f}\sinh(\b \t+\f), \quad
v=\frac{e^{E\t}}{\cosh\f}\sinh(\b\t-\f), 
\ee
which bear a remarkable resemblance to the time-like geodesics the
difference now being, $\tanh^2\f=\frac{E^2}{\b^2}$). These satisfy
$uv<\tanh^2\f<1$, and hence do not reach the singularity, but extend
across both regions I and II of the black hole geometry, similar to
the timelike geodesics (see the magenta curve in
Fig:\ref{timelike-geodesics}).




For this solution, we can calculate the Kac-Moody charges:
\bea
\tilde J^{(2)}_\pm = -\frac{k\b\tanh\f}{2} \,&;&
\,\, \tilde T_\pm= \frac{k}{4} \b^2, \\\nn
J^{(2)}_\pm
=\frac{k}{2}(E-\b\tanh\f)=0\,&;&\,\,
T_\pm=\frac{k}{4}(\b^2+E^2-2E\b\tanh\f)
\eea
In this case, we again get $\tilde T_\pm>0$, as expected from the
continuous series. But since $|J^{(2)} _\pm|<|\b|$, these form a
double sided representation of \sl2.



It is also clear from the above that the spacelike geodesics of the
black hole geometry are obtained from the spacelike geodesics of \sl2
in the same manner as the null and the massive geodesics.


\section{Regions V and VI: `Behind' the singularity}

In view of the observation that the physical geodesics constructed in
the previous sections do not extend beyond $uv=1$ into the region
``behind'' the singularity, it is of interest to examine the nature of
the geodesics in this region.


\subsection*{Timelike geodesics}

The ones with energy $E^2>m^2$ are given by
\be
u=\frac{1}{\sinh\f} e^{-E\t}\cosh(\b\t-\f), \quad
v=\frac{1}{\sinh\f} e^{E\t}\cosh(\b\t+\f)
\ee
where $\tanh^2\f=\frac{\b^2}{E^2}$. Using the above, we see that these
reach infinity - i.e., represent scattering states as before. But,
more interestingly, we get 
$uv=\frac{\sinh^2\b\t+\cosh^2\f}{\sinh^2\f}>1$, i.e, the geodesics
never reach the singularity at $uv=1$.  This supports the idea that
the geodesics in this region are independent of those in the regions
``in front'' of the singularity (as suggested by the spacelike
geodesics of the previous section).

Again, we can find matrices
\bea
\tilde g^V _+(\sp)  &=& \begin{pmatrix}\cosh(\b \sp/2-2\f) &\cosh(\b \sp/2+\f) \cr
\cosh(\b \sp/2-\f) &\cosh(\b \sp/2+2\f)\end{pmatrix}, 
\\\nn
\tilde g^V _-(\sm)&=&\begin{pmatrix}\sinh(\b \sm/2-\f) &-\sinh(\b \sm/2-2\f) \cr
-\sinh(\b \sm/2+2\f) &\cosh(\b \sm/2+\f)\end{pmatrix},
\eea
which give rise to the geodesic above after spectral flow.  The various quantum numbers for
these timelike geodesics turn out to be
\bea
\tilde T_{++}&=&\tilde T_{--}=\frac{k}{4}\b^2; \quad \tilde J_{\pm}=\frac{k}{2}\b\coth\f,\\
T_{++}&=&T_{--}=\frac{k}{4}(\b^2+E^2+2\b E\coth\f); \quad J_{+} ^2=J_-^2=\frac{k}{2} (E+\b\coth\f),
\eea
implying that these also belong to the principal continuous series.
And as before, these are obtained by transforming spacelike geodesics in \sl2 with
\[U=\left(\barr 
e^{\frac{-\f}{2}} & -e^{\frac{\f}{2}}  \\
e^{-\frac{\f}{2}}  & -e^{\frac{-\f}{2}}  \earr 
\right) {\rm and }\qquad
V=\left(\barr 
-e^{\frac{\f}{2}}  & e^{\frac{-\f}{2}} \\
e^{\frac{-\f}{2}}  & -e^{\frac{\f}{2}} \earr\right).\]

\subsection*{Null geodesics}
The null geodesics in the region $uv\geq 1$ are
\bea
u=-k; \,\,\,\, v=-(k e^{+2E\t}+\frac{1}{k})\\\nn
v=-k; \,\,\,\, u=-(\frac{1}{k}+k e^{-2E\t})\nn
\eea
and are distinguished from those in region I by the relative plus sign
between the $e^{2E\t}$ and the $\frac{1}{k}$ terms. Consider a particular null geodesic
\bea
v=-(1+e^{2\b\t}), \quad u=-1,
\eea 
where the constants are chosen such that $\t \to -\infty$ corresponds
to the singularity $uv=1$.  Observe that the derivatives
$\frac{d(uv)}{d\t}$ and $\frac{d(u/v)}{d\t}$ both vanish as we tend to
the singularity. On the other side of the singularity, the null
geodesic in region II ($uv<1$) with the same quantum numbers given
by \be v=(-1+e^{-2\b\t}), \quad u=-1, \ee also has these derivatives
vanishing as we approach the singularity (in the future).  Thus, these
two geodesics cannot be argued to be the continuation of each other
through the singularity.

\subsection*{Bound states}

It is easy to show that there are {\em no} solutions with $E^2<m^2$ in this
region of the geometry. The localised geodesics belonging to the
discrete series live only in the regions in ``front'' of the
singularity.

\section{Building up representations \label{reps}}

To summarise the results of the previous sections, observations of the
timelike geodesics in regions V and VI of the black hole geometry
suggest that these are independent states from the corresponding
geodesics in regions I and II.  The spacelike geodesics of these
regions also suggest a similar conclusion.

The timelike geodesics of region I continue across to region 
II - but are discontinuous at the singularity (see the dashed curves
in Fig. \ref{timelike-geodesics}).  However, we observed that like the
spacelike and timelike geodesics with $E^2>m^2$, the null geodesics
are also obtained from spacelike geodesics in \sl2. Thus, we conclude
that null geodesics of region I must also be paired with a
corresponding null geodesic of region II.  Also, as argued in the
previous section, this `pair' is independent of the null geodesic in
regions V and VI. A similar `pairing' must occur for the spacelike
geodesics as well. 

Thus, in the quantum theory we expect two copies of states with
parameters $|E,\b,J^2\rangle$ - one representing the states in regions
I and II, and the other representing the pair in regions V and VI.
On the other hand, the absence of the timelike geodesics with
$E^2<m^2$ in regions V and VI implies that a single copy of states for
these geodesics suffices.
These observations are borne out by the representation theory of \sl2
\cite{Mukunda} as well.

By using the states of the quantum theory, we can construct
wavepackets which ``follow'' the geodesic (coherent states). 
In this construction, the classical initial conditions will appear as
parameters of the wavefunctions.  Symmetry transformations acting on
the classical solutions can be interpreted as changing the initial
conditions. This action must therefore lift to an action on the
wavefunctions as well. Hence, if there is some set of initial
conditions which is closed under the action of the (symmetry) group -
then this set will form a representation of the (symmetry) group upon
quantization.  We can therefore try to understand how the
representation is filled out by examining the classical geodesics.

We first note that the trace of a matrix in \sl2 is a conjugacy class
invariant. Therefore, regions V and VI which correspond to $uv>0$ form
the conjugacy classes $|Tr(g)|=|a+b|<2$. Hence, under the action of
the vectorial symmetry (which is conjugation), the trajectory now
viewed as an \sl2 matrix cannot be moved into the other regions of the
black hole geometry. If we regard the vectorial action as changing the
initial conditions of the geodesic, then it follows that all points on
the geodesics in region V and VI form a closed set of initial
conditions (of the vectorial action).

	

Writing the classical solutions as \sl2 matrices, we can easily see
that only the following \sl2 $\times$ \sl2 transformations commute with the spectral flow, 
\bea
g\to e^{\s_3 \l} g e^{-\s_3 \l}, \\ \nn
g \to -g, \\ \nn
g\to i\s_2\, g\, i\s_2,
\eea
and compositions of these transformations
These, when applied to a geodesic produce another classical geodesic
of the black hole. 
The first shifts $(u,v)\to (u \l^2, \frac{v}{\l^2})$ - which preserves
the hyperbola $uv={\rm constant}$.  The second changes the sign of
$u,v$. Both produce new classical solutions with different initial
conditions while keeping the quantum numbers $E,\b,J^{(2)}$ fixed.  The
third transformation interchanges $(u,v) \to (-v,-u)$ and results in
flipping the signs of $E,\b$. Thus, it maps the outgoing (from the
singularity) solution in region I ($u>0,v<0$) to different infalling
trajectory in region II ($u<0,v>0$) and vice versa.

The most general solution of the geodesic equations will have four
integration constants. Two of those can be taken to be $E$ and
$\b$. The third parameter is the constant $\l$ in the first of the
transformations above - this shifts the origin of Schwarzschild
time. The last constant is the parameter $\t_0$ which shifts the
affine parameter $\t$ of the geodesics. Thus, once we have considered
the action of the above transformations, we have accounted for all
geodesics.

It should be noted that both $\b>0$ and $\b<0$ give rise to the same
spacelike or timelike ($E^2>m^2$) geodesics (changing sign of $\b$
also changes the sign of $\f$) provided we consider the entire
trajectory. For null geodesics, this means we consider the pair
together.  In particular, the signs of both $\frac{dt}{d\t}$ and
$\frac{d\r}{d\t}$ are not affected by $\b\to -\b$.  In view of this,
it is necessary to retain only $\b>0$ states in the spectrum of the
theory (or $\b<0$, of course) to account for all the time like
geodesics of the black hole geometry.

We can put together the above observations thus; for each set of
quantum numbers, we have two states representing the pairs in regions
I and II and V and VI respectively, 
\be \mathcal{C}_\b ^E = \{ \left(\begin{array}{cc} |+E,\b,+\l \rangle \\ |-E,\b,-\l\rangle\end{array}\right), \quad E, \l\in \mathbb{R}, \quad \b>0   \}.\ee
In this case, the $J^{(2)}$ operator will act as $ \l {\bf \s_3}$
where $\l$ is the $J^{(2)}$ eigenvalue.  This is the doubled structure
of the continuous series representations of $so(2,1)$ argued for in
\cite{Mukunda}.  In that work, it was argued that for the global time
\sl2 translations to be properly represented, the generator $J^{a}$
should be represented $2\times2$ matrices along with being
differential operators.

The above representation of the state is consistent with the
observation that if one part of the paired timelike or null geodesics
are infalling in region I (V), then the other is outgoing in region II
(VI). We also note that the sign of $E$ is correlated with that of
$J^{(2)}$ via spectral flow.

Under time reversal the ket $|E,\b,J^{(2)}\rangle$, is mapped to
$|-E,\b,-J^{(2)}\rangle$ and thus the above doubled state is mapped to
itself (upto a phase)\footnote{Time reversal should be represented as
  an anti-unitary operator, we consider only the unitary part of such
  an operator}.  Time reversal transformation generated by $i\s_2$ is
equivalent to translation by $\pi$ along the (compact) global time
direction in \sl2 ($AdS_3$) which is generated by $J^{(0)}$ (this is
easily seen using global $AdS_3$ charts as given in, say, \cite{mog}).
Thus, $J^{(0)}$ operator of \sl2 acts as $J^{(0)}\otimes \mathbb{I}$.
This structure of the \sl2 (or current algebra) operators is also
consistent with the assignment in \cite{Mukunda}.

All in all, we conclude that geodesics of the extended geometry fall
into three representations of \sl2: the discrete series $D_j ^E$ and
the continuous series $C_\b ^E$ with $\b >0$.  In each of these
representations, the $J^{(2)}$ eigenvalue occurs twice and is
understood as labelling the states in regions I,II and V,VI
respectively.
The $D_j ^E$ on the other hand does not have such a doubling, and
represents the localised geodesics, which are {\em not} present in
regions V and VI.

\section{Vector-Axial duality and new states \label{VAD}}

It is well known that gauging either the axial action (which is what
we have been considering until now) or the vectorial action of the
diagonal $R$-subgroup in \sl2$\times$\sl2, one obtains the same
(extended) target space. This is the analog of the spacetime
operation $R\to\frac{1}{R}$  in the usual T-duality of string theory.

Under the vectorial gauging action, the $a,b$ entries of the \sl2 matrix are
invariant while $(u,v)$ transform to $ (\L^{2} u, \L^{-2}v)$ respectively.  Thus, in this case,
the black hole target space is described using the $a,b$ coordinates
of \sl2. Since $ab=1-uv$, the asymptotically flat region in ``front''
of the horizon $uv<0$ is mapped to the region ``behind'' the
singularity $ab=1-uv>1$.
Hence, we can interpret the entries $a(\t,\s)$ and $b(\t,\s)$ from all
the \sl2 matrices $\left(\begin{array}{cc} a & u\\-v&b \end{array}
\right)$ representing
the worldsheet solutions of the previous sections, as describing new sigma model solutions in the dual region
of the black hole spacetime (in the $u,v$ coordinate chart).  From the
original point of view (i.e., axial gauging), this is equivalent to a
right (or left) multiplication of the \sl2 matrix worldsheet by
$\left(\begin{array}{cc} 0 & 1\\-1 & 0 \end{array} \right)$ and hence
gives another solution of the sigma model equations.  Clearly, this
operation results in $J^{(2)}_\pm\to \pm J^{(2)}_{\pm}$.

We may alternately, not only perform the target space transformation
above, but supplement it with a worldsheet $\t \to \s$ interchange
This operation is analogous to flipping the sign of the right moving
momentum in the usual circle T-duality considerations.  If this
T-duality is a symmetry of the full string theory, then this will map
solutions to solutions (in the textbook example of T-duality on a
circle, this sends the state $|(n,w)\rangle\to |(w,n)\rangle$)
In our case, this will produce a new worldsheet solution.

For instance, in the matrix given by  Eq. (\ref{null-lift}), representing a null
geodesic in region I, right multiplication by $i\s_2$ gives
$$u=\exp(E(\t-\s)),\quad v=\exp(E(\t+\s)),$$ from which, after
interchanging $\t\to\s$, we get
$$uv=\cosh^2\r=e^{2E\s}, \quad t=-E\t,$$ which we could expect to be a
solution in region V (provided the Virasoro conditions are met, of
course). On the other hand,  we start with the matrix representing a null geodesic in
region V, we obtain a string worldsheet that covers all of regions I and II
$$-uv=\sinh^2 \r= e^{-E\s} \quad t=-E\t.$$
Motivated by the AdS/CFT correspondence, we propose that such
worldsheets extending to the boundary should be interpreted as
operators of the boundary theory.

\begin{figure}
  \begin{center}
  \includegraphics[height=5cm]{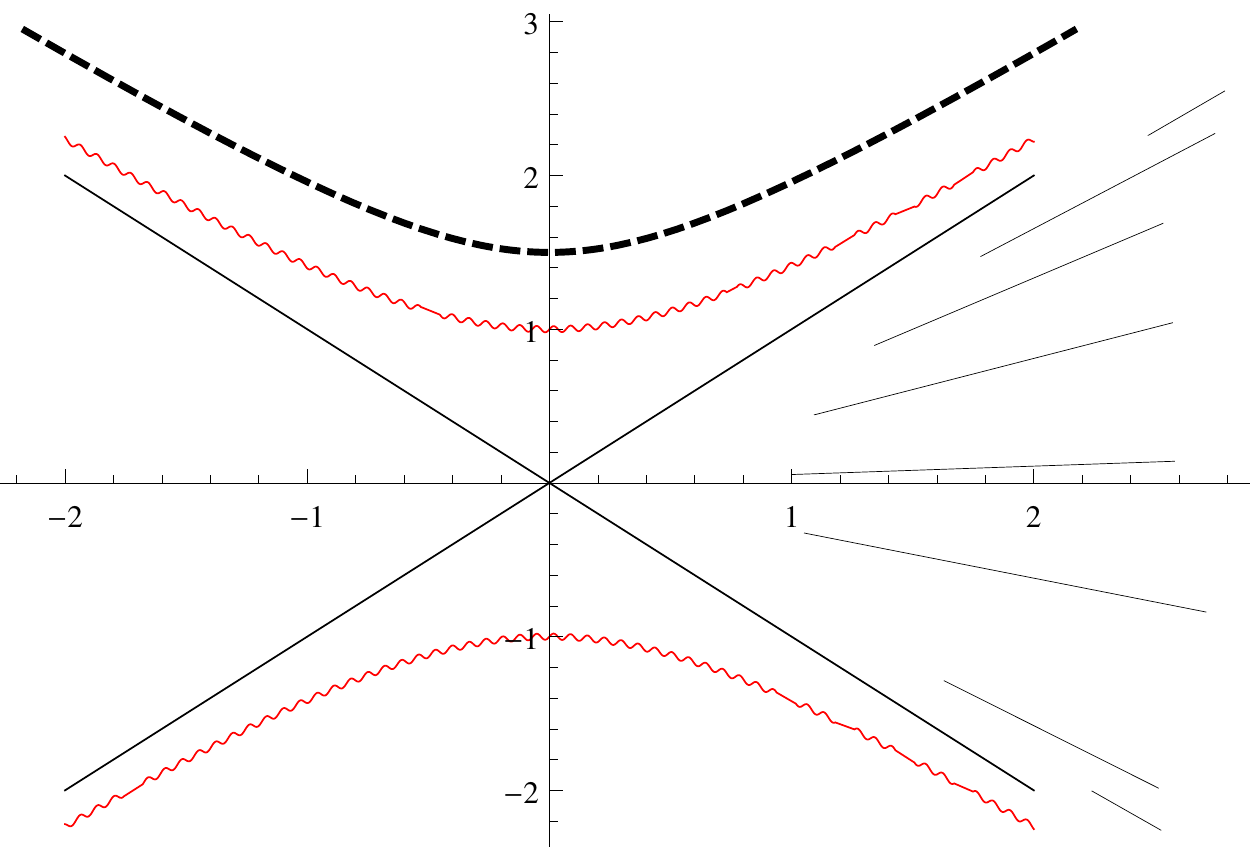}
         \caption[Example of figure]{The new stringy solutions}%
         \label{newstrings}
\end{center}
  \end{figure}
Upon dualising, the timelike geodesics of region V (the black dashed
curve in Fig:\ref{newstrings}) give worldsheets
$$uv=-\frac{\cosh^2\b\s}{\sinh^2\f},$$ which extend out to the
boundary in the asymptotically flat regions I \& II, but do not reach
the horizon. These worldsheets, for a few instants of time, are shown
as black segments in the right quadrant in Fig:\ref{newstrings}.
These ``folded strings'' are similar to the long strings considered in
\cite{Mal} in that the tip scatters out from the region near the
horizon. These worldsheets are naturally better interpreted as (single
trace) operators of the boundary theory.

The implication of this duality is that for a given set of quantum numbers
$E,\b,J^{(2)}$ there are two solutions in each asymptotically flat
region. We have the worldsheet representing the geodesic in this
region, or the solution obtained by dualising the corresponding
geodesic in region V which we can choose to have the opposite sign for
$\b$.  Note that the solution and its `dual' version do not exist
simultaneously in the same asymptotically flat regions (I,II) or (V,
VI).

Using the earlier assignment of quantum numbers for the geodesics, we
might write these as
\Bigg(\begin{tabular}{cc} $|+E, -\b,  \l \rangle$ \\ $|+E,-\b, \l\rangle $\end{tabular}\Bigg)
with $\b>0$. 
This assignment of quantum numbers is natural 
in terms of the Seiberg-bound states and Seiberg anti-bound operators
of the matrix model (which are indeed constructed by $j\to1-j$ in the
quantum theory - $j=\frac{1}{2}-i\b$).

\subsection{Horizon Strings}

We reserve the most interesting case for the last - the geodesics
with $E^2<m^2$ (belonging to the discrete series). In this case, the
dual solution is $$uv=\frac{\sin^2\b\s}{\sin^2\f}$$ which is always
inside the horizon but extends across the singularity ($0\leq uv \leq
{\rm cosec}^2\f$).
\begin{figure}
  \begin{center}
  \includegraphics[height=5cm]{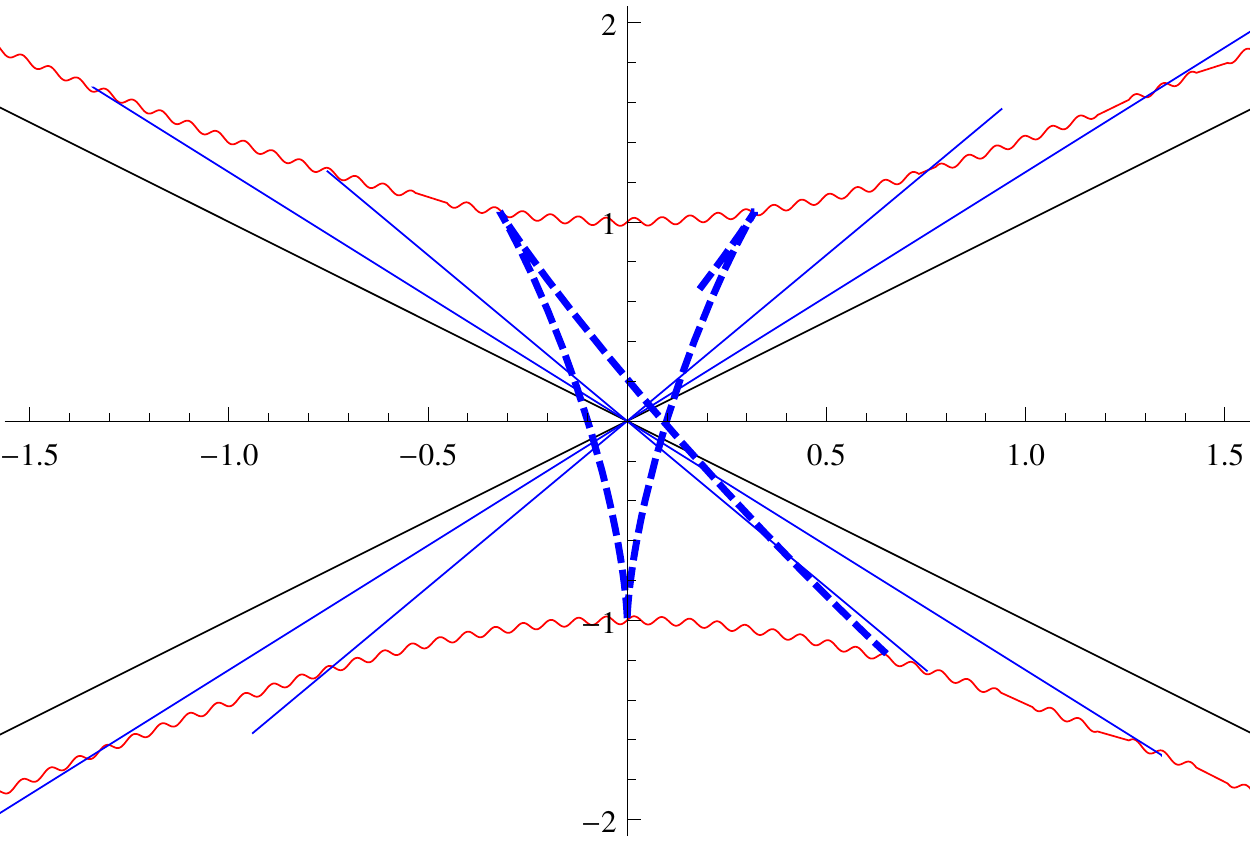}
         \caption[Example of figure]{The near horizon worldsheets}%
         \label{horizonstrings}
         \end{center}
\end{figure}

These worldsheets do not extend to infinity and since we might expect
that $\b$ is an integer, are doubly folded over at $\s=0,2\pi$ and
thus meson-like These string worldsheets (for a few instants of time)
are shown as thin (blue) lines in Fig:\ref{horizonstrings} that extend
a little across the singularity (which is the red wavy hyperbola).

Thus, we might choose to regard the black hole as a condensate of
``mesons'' (to use AdS/CFT terminology) consistent with a holographic
interpretation of the finite temperature state of the dual theory.
Since these folded mesonic strings are visible only in regions V and
VI, this interpretation is valid only for the boundary observer in
this region(s).

Or else, the black hole can be considered as a bound state, with the
localised geodesics being the internal degrees of freedom visible only
to the external observer in regions I and II (in Fig:
\ref{horizonstrings}, these are the dashed curves in the middle).

These are complementary to each other in the sense of vector-axial
duality and also possibly complementary to each other in the sense of
black hole complementarity.  However, only in the region between the
horizon and the boundary, we have both descriptions as being
simultaneously applicable.

It should be emphasized that these are complementary (in $\a'$)
descriptions of the Hilbert space of states interior to the black hole
and assumes that the vector-axial duality is a symmetry of the string
theory.

\section{Analysis of the spectrum \label{spectr}}

The conditions for a classical solution to be a physical string configuration
is
\be
T^{tot}_{++}=T_{++} ^{WZW} -\frac{(J^{(2)})^2}{k}+T^I _{++}= 0, \qquad J_+  ^{(2)}= J_-^{(2)}. \label{cond}
\ee

The first condition is, of course, the Virasoro condition for the total
stress tensor where we have included an ``internal'' CFT with stress
tensor $T^I$.  A similar condition applies for the right moving
$T_{--}$ as well - in what follows, we shall focus exclusively on the
left moving currents. The full spectrum of the string theory is
obtained by putting together the left and right moving states so that
they satisfy the level matching condition given above.

We observer that, for any solution $g$, if $\tilde T- \frac{1}{k}(\tilde J^{(2)})^2=0$, then the
solutions obtained by spectral flow, Eq. \ref{T-flow},  along the $J^{(2)}$ direction
satisfy
\[T-\frac{1}{k} (J^{(2)})^2=  \tilde T + E \tilde J^{(2)} +\frac{k}{4}E^2-\frac{1}{k} (\tilde J^{(2)}+\frac{k}{2} E)^2=0 \]
and thus give rise to a family of solutions labelled by $E$.  

The second condition on the currents $J_\pm ^{(2)} $ in Eq. \ref{cond}
comes from the gauging and is interpretable as a level matching
condition. The spectral flow acts the same way on $J^{(2)} _\pm$ and
hence, if this level matching condition is satisfied before spectral flow, it
will be automatic afterwards.


\subsection{Null Geodesics}

We first point out that the matrix $\tilde g$ of Eq. \ref{gtn} that
represents a spacelike geodesic in $AdS_3$ itself satisfies the
physical state conditions of the coset model.  However, it does not
represent a fixed energy trajectory of the black hole.  That is to
say, $\tanh^2\r \frac{dt}{d\t}$ is not a c-number.
Yet, this configuration is an eigenstate of $J^{(2)}$ (that is
$J^{(2)}$ is a c-number). This is an additional reason why $J^{(2)}$
must not be identified with the spacetime energy of the particle.




From the expressions given in Eq. \ref{T-null}, we can see that the
Virasoro constraints of the gauged theory are satisfied for the matrix
$g$ for any $E,\b$. 
Thus, these are physical solutions of the black hole sigma
model. These equations, not unexpectedly, resemble the dispersion
relation for massless particles since it is known that there are
massless particles in the spectrum of this theory. The important
difference is that we have massless particles for every pair $(\b,E)$.

Observing that $T_{++}$ of the WZW is bounded below, i.e.,
$T_{++}>0$ ({\it before} ``spectral flow'') allows to infer that these
particles must be in the principal continuous representation of \sl2
(for principal continuous series the quadratic Casimir of \sl2 is
bounded below by $\frac{1}{4}+\b^2$).

In the case the string theory includes an `internal' CFT - the
massless states above continue to be physical, provided the internal
CFT has states with $T^I _{++}=h=0$ (since the condition is $T_{++}+T^I
_{++}=0$).

\subsection{Timelike geodesics}

For  $E^2>m^2$, the physical state condition is 
\be \tilde T^{tot}={\tilde T}^{SL_2}-\frac{1}{k} ({\tilde J}^2)^2+h = \frac{-k\b^2}{4} {\rm cosech}^2\f+h=0. \ee
Thus, these massive geodesics can be physical in the {\em classical}
string theory provided the internal CFT contributes a positive weight
$h$.
Similarly, for the geodesics with $E^2<m^2$, we see that the total
stress tensor is $$ {\tilde T}^{tot} _\pm=-\frac{k\b^2\sec^2\f}{4}+h=0,$$ which can be
satisfied if $h>0$.
The situation corresponding to $\b=0$ is the case when we start with a
constant matrix in \sl2 and then perform a spectral flow on it 
\be
g=\left(\begin{array}{cc}
a\,e^{E\sigma}& u\, e^{-E\t} \\
-v\,e^{E\tau}&e^{E(-\sigma)}\earr\right).
\ee
These solutions of the sigma model equations of motion correspond to
$uv={\rm const}$. The existence of such solutions is analogous to similar
solutions in \cite{mog}. 
As for the spacelike geodesics, these are never physical because 
\be {\tilde T }^{tot}={\tilde T}^{SL_2}-\frac{1}{k} ({\tilde J}^2)^2+h = \frac{k\b^2}{4} - \frac{k}{4} \b^2 \tanh^2\f+h=\frac{k \b^2}{4}{\rm cosech}^2\f+h>0, \ee
as long as the ``internal'' CFT has $h>0$.

\subsection{New string solutions}

The new string solutions are constructed from \sl2 solutions, by right
multiplying by $i\s_2$ and then interchanging $\t,\s$. Under each
operation the right moving currents $J^a _-$ pick up a negative sign
and hence remain unchanged in the end. Thus, the analysis of the
previous section can be carried over in toto and the new solutions are
valid classical solutions if the old ones are.

\subsection{Quantum analysis}

The space of states of a coset conformal field theory is normally
constructed by starting with representations of the parent CFT and
writing the states in a basis adapted to the action of the quotienting
subgroup. In this adapted basis, it is easy to impose the gauging
conditions, and one can identify the states of the coset sigma model.

In the quantum theory, the physical state condition for an axially gauged coset model can be written \cite{Karabali, Gawedzki}
\be
L_0=L_0 ^{(SL_2)}-\frac{1}{k} (J_0 ^{(2)} )^2=1,\qquad J^{(2)}_m=\bar J^{(2)} _m=0,  \forall\, m>0,  \qquad J_0 ^{(2)}+\bar J ^{(2)} _0=0,\label{physicalstate}
\ee
where, for \sl2
\be
{\tilde L}_0=\frac{-j(j-1)}{k-2} +N, \hspace{1cm} {\tilde J}^{(2)}_0=\l
\ee
with $N$ being the Kac-Moody level. The third conditions arises from
the gauging procedure.  The $J^{(2)}_0$ quantum number $\l$ is
unrelated to the Casimir (although in the Discrete series representations
there are restrictions on the values of $J^{(2)}_0$ depending on $j$).

For the continuous series representations of \sl2, we have $j=\frac{1}{2}+i
\b$, so that for the null and timelike geodesics with $E^2>m^2$, we
get
$$ \frac{\frac{1}{4}+\b^2}{k-2}+N+h - \frac{\l^2}{k} =1$$ as the
on-shell condition. If we assume that $N=0$, that there is
no ``internal'' CFT and set $k=\frac{9}{4}$, this reduces to the
dispersion relation of the usual `tachyon' of the critical (c=26) 2D
black hole $\b^2=\frac{4}{9} \l^2$ .
However, we have a subtlety
\cite{Mukunda}.  The states of the nonexceptional continuous series of
representations, when written in terms of the eigenstates of $J^{(2)}
_0$ are 'doubled'. That is to say, a single irrep of \sl2 requires
that each `momentum state' $J^{(2)} _0 |\l \rangle=\l |\l\rangle$ appear
twice - so that a basis for this Hilbert space takes the form of a column vector
\be {\mathcal C}_\b= {\rm Span} \{ \left( \begin{array}{cc} |\b,+\l \rangle \\  |\b,-\l \rangle \end{array}\right) , \l \in \mathbb{R},\b>0\} \ee
On these vectors, the Hermitean $J^{(2)}_0$ operator acts as
$J^{(2)}_0=\l \otimes \s_3$. The other generators are also now written as $2\times2$
matrices - so that the \sl2 commutation relations are satisfied.

This fits nicely with the conclusions drawn in section \ref{reps} from
studying the geodesics which led to the suggestion that timelike and
null geodesics in regions I and II are paired as a single state. The
second copy then refers to regions V and VI.  The difference in sign
can be related to the observation that if a timelike (or null)
geodesic is infalling in region I, then its partner in region II is
outgoing. We have also remarked that we shall choose $\b>0$ for region
I and $\b<0$ for region V (and similarly for regions II and VI). This
differs slightly from the suggestion made in \cite{Mukunda} that the
doubling of $J^{(2)}$ could be related to the option of choosing
either sign for $b$.

For the discrete series, we know that $\frac{1}{2}<j<\frac{k-1}{2}$
and hence the mass shell condition at level zero is
$$\frac{-j (j-1)}{k-2}+N+h - \frac{\l^2}{k} =1.$$ These states exist
for special values of the momentum \cite{DistNel} as determined by
this condition, but the energy of these states is given by the
spectral flow parameter and can take any value. This is possible
because in the hyperbolic basis the eigenvalues of $J^{(2)}$ can also
be pure imaginary in a hermitean representation.

Given the above matrix form of the zero modes of the KM generators
$J^{(a)}_0 $, we need to determine the form of the remaining modes $J^a _m$ of
the currents so that the commutation relations 
\bea
    [ J^{2} _m, J^{2}  _n ]=\frac{k}{2} n \d_{m+n,0}, \quad
    [ J^{(2)}_n, J^\pm _m ]=\pm i J_{n+m} ^\pm, \quad
    [J^+ _n, J^{-} _{m} ]=-2i J^{(2)} _{m+n} - k n \d_{m+n,0} ,
\eea
of the current algebra are preserved in the continuous series
representation ${\mathcal C}_\b$. The simplest assumption is that the
$J^a _m$ are tensored with the same $2\times2$ matrix as the
corresponding $J^a _0$. This assumption is consistent with the
spectral flow operation and is clearly also consistent with the
physical state conditions Eq. \ref{physicalstate} of the coset sigma model
\cite{Gawedzki}.

Bringing in spectral flow into the discussion does not alter the above
conclusions about the mass shell conditions. This is because, as
remarked above, if a state satisfies the mass shell condition before
spectral flow, it will satisfy it after the spectral flow as well.

\section{Summary and Discussion \label{summary}}

To summarise the results presented, we have seen that the study of the
action of symmetries on the space of geodesics has proven to be highly
profitable.
We have been able to cast several features of the representation
theory of \sl2 into properties exhibited by the geodesics.
The doubling of the representations of the $J^{(2)}, J^+$ algebra to form
a single irreducible represention of \sl2 (for the Continuous
series) was motivated in \cite{Mukunda} as a
requirement that the generator of global time translations be properly
represented. Here, we find that it is better motivated as the
requirement that time reversal be representable.

This doubling of the $J^{(2)},J^+$ representation is absent for the
Discrete series (these are absent in regions V, VI).  These are
localized in the near horizon region, and are captured by the square
integrable wavefunctions making up the Discrete series.  The geodesic
analysis clearly leaves open the physical interpretation of the
eigenvalue of $J^{(2)}$ maybe along the lines of the phase shift
studied in \cite{natsuume-pol}.

A primary issue is to understand whether the operation of ``spectral
flow'' along the hyperbolic direction makes sense in the CFT. While we
have argued for it classically, there does not appear to be any
problem at a formal level in generalizing to the quantum theory.  In
fact, to obtain all possible values of $E$, we must allow all values
for the spectral flow parameter $E$.  A separate question is whether
these flowed states are necessary.  However, we can expect that the
gravitational backreaction of any finite energy particle in 1+1-d will
lead to large deformations of the asymptotic region. Thus, to preserve
asymptotics we might impose the condition that $E=0$.

We have also exhibited several new features of the states of this
sigma model.  The states arising from the discrete series play the
role of the winding strings of the Euclidean `Cigar' theory.  However,
these are only visible to one (pair) of the asymptotic regions.  The
winding strings of the T-dual `Trumpet' geometry are mapped to the
localized worldsheets visible only to regions V and VI (which form the
asymptotics of the Trumpet geometry).  This complementary view of the
`interior' is clearly a stringy effect and is worth exploring further.
Similarly, the states of the continuous series have two descriptions -
either as scattering states or as worldsheets ending at the boundary
(and thus operators) and involving a $j \to 1-j$ flip suggesting that
this is a state-operator duality. The folded worldsheets ending at the
boundary could be related to the additional non-singlet degrees of
freedom used by \cite{betzios} to obtain the KKK model. This suggests
comparing correlation functions obtained from this matrix model and
that of the strings described using ideas of holography.

These observations are pertinent if we wish to regard the 2D black
hole as the high temperature phase of a boundary gauge theory (the
matrix model). The geodesic analysis suggests that we should regard
the boundary as two points - the asymptotics of region I and II (or
dually regions V and VI). It has been suggested \cite{Witten2} that
region II of the black hole geometry is mapped to the second
asymptotic of the potential in the matrix model. This also suggests
that we might search for a Hawking-Page type phase transition to a low
temperature (linear dilaton) phase.

A related question is the issue of conserved charges carried by the
various strings. In \cite{Sen}, the $W_\infty$ charge carried by the
2D black hole was computed. If the horizon strings are indeed the
degrees of freedom, then they should also carry some part of the same
$W_\infty$ charges.


It may be facile to expect that the special features noted above will
extend in a simple manner to other dimensions. However, at least the
existence of strings ending on the horizon, and their dual relation to
localized geodesics seems generalizable \cite{Doran}.

One interesting question is to construct the characters of these
representations and hence forming the partition function.  One could
further compare with the partition function of the Euclidean black
hole \cite{Part} and trace the winding modes to the Lorentzian
geometry. In this regard, a significant contribution to answering this
question already appears in the work of \cite{Israel}.  In this work,
the authors construct the partition function for Lorentzian $AdS_3$ by
building upon characters of the gauged \sl2 theory.  Using this, they
have also constructed partition functions for various marginal
deformations - in particular, the Lorentzian black hole (see eqns 5.15
and 5.17).  It will be of much interest to read off the spectrum of
states from the partition function.  We should compare the structure
of the spectrum with the work of \cite{Giveon-Tr}. In this work, the
(worldsheet) elliptic genus of the Euclidean black hole which included
states from the discrete series was argued to satisfy a curious
identity. This identity should be related to the observation that
while the timelike geodesics with $E^2<m^2$ are absent in the region
V,VI of the black hole, one can construct ``T-dual'' string
configurations which satisfy the physical state conditions.  The
computation of various correlation functions - either from the point
of view of the regions I and II or from the T-dual regions V and VI is
another question. As in AdS holography, the geodesics and evaluation
of the action can be used to obtain a saddle point approximation to
correlation functions.



One can use similar techniques to study the third gauging of \sl2
(i.e., gauging the lightcone (parabolic) direction) which gives rise to
Liouville theory. We can again ask whether the spectrally flowed
representations of \sl2 survive in the coset theory. In this context,
Balog et. al., \cite{Balog} have found topological sectors from a study
of the Virasoro co-adjoint orbits in the case of the Liouville theory.

\section*{Acknowledgements}

I am grateful to Pravabati Chingangbam for a critical reading of the
manuscript and motivation. I also acknowledge useful conversations
with Alok Maharana, Kapil Paranjape and C. G. Venketasubramanian.  I
would like to thank HRI, Allahabad for their warm hospitality and a
motivating atmosphere where some of this work was completed,

\section{Appendices}
\subsection{Conventions}
In this appendix, we present the details of the gauged sigma model to make our conventions clear. The action for the \sl2 CFT is given by 
\bea
S_{WZNW}=\frac{k}{8\pi}\int d^2\sigma\sqrt{-h}
Tr\left(\del_a\,g\del^a\,g^{-1}\right) + k\, \Gamma \\\nn
\Gamma=\frac{1}{12\pi}\int_B \epsilon^{abc}
Tr(\del_ag\,g^{-1}\del_bg\,g^{-1}\del_cg\,g^{-1})
\eea
and the trace is calculated in the two dimensional representation of SL(2,R) and $\alpha'=1$.

Parametrizing $g=e^{t_L\sigma_3}\,e^{\rho\sigma_1}\,e^{t_R\sigma_3}$, 
the kinetic term gives
\be
L=-\frac{k}{4\pi}\int d^2\sigma\sqrt{-h}
\left(\del_\alpha\rho\del^\alpha\rho+\del_\alpha t_L\del^\alpha
t_L+\del_\alpha t_R\del^\alpha t_R + 2\del_\alpha t_R \del^\alpha t_L
\cosh2\rho\right),
\ee
and the WZ term gives
\be
\Gamma=-\frac{k}{2\pi}\int d^2\sigma\cosh2\rho
\epsilon^{\alpha\beta}\del_\alpha t_L \del_\beta t_R.
\ee
From the kinetic term above, one can read off the metric of the WZW
model which turns out to be
\be
ds^2=k(d\r^2+ \cosh^2\r d\f^2-\sinh^2\r dt^2)\,\,\,\,\ t_{R,L}=\frac{\f\pm t}{2}
\ee
By contrast the metric in global \sl2 (or $AdS_3$) co-ordinates is
$ds^2=k(d\r^2-\cosh^2\r dt^2+\sinh^2\r d\f^2)$.
 
We chose the normalization of the of the \sl2 generators such that
$Tr(\tau^a \tau^b)=\frac{1}{2}\eta^{ab}$ and where
$\eta_{ab}=diag(-1,1,1)$. More specifically,
$\tau^0=\frac{i\sigma^2}{2}, \tau^1=\frac{\sigma^1}{2},
\tau^2=\frac{\sigma^3}{2}$. Note that  $\tau^0$ is not Hermitean.
The conserved currents of the WZNW model are defined as
$J_+=-k(\del_+g\,g^{-1})$ and $J_-=k(g^{-1}\,\del_-g)$ and the components are
defined by $J^a_\pm= Tr(\tau^a J_\pm)$.
and the mode expansions are 
\be
J_L ^a=\sum_n J^a _n e^{-i n \s^-}\qquad 
J_R ^a=\sum_n \tilde J ^a _n e^{-i n \s^+}
\ee
The stress tensor ({\it defined} as
$T_{ab}=\frac{-4\pi}{\sqrt{-h}}\frac{\delta S}{\delta h^{ab}}$) turns out to
be
$T_{++}=\frac{1}{k}\eta_{ab}J^a_+ J^b_+$ in terms of the current.

\subsection{Gauging}
Adding the two terms in the action ($\epsilon_{01}=1$), and rewriting in $\s_\pm$-variables on the
worldsheet, we get
\be
L=\frac{k}{2\pi}\left(\del_+\rho\del_-\rho+\del_- t_L\del_+t_L
+\del_+ t_R\del_- t_R + 2\del_+ t_R \del_-t_L \cosh2\rho\right)
\ee
where $\del_\pm=\ha(\del_\t\pm\del_\s)$. The black hole sigma model is obtained by   
gauging a simultaneous translation of  $t_{L,R}$. This is accomplished by adding the
extra terms involving gauge fields $A_\pm$ 
\bea\nn
L_{gauge}=\frac{k}{\pi}\left(A_+(\del_-t_R+\cosh2\rho \del_-t_L)
+A_-(\cosh2\rho\del_+t_R+\del_+t_L) -A_-A_+(1+\cosh2\rho)\right)\\\nn
=\frac{k}{\pi}\left(
\frac{1}{k}(A_+ J_-^{(2)} - A_-J_+^{(2)}) - A_-A_+(1+\cosh2\rho)\right)\hspace{1.5cm}
\eea
and prescribing the following gauge transformation properties for $A_\pm$
$$\d t_{L,R}=\L \qquad \d A_\pm=\del_\pm \L$$.
Solving the e.o.m for the gauge fields, we get
\be
A_+=\frac{(\delp t_L+\cosh 2\r \delp t_R)}{(1+\cosh 2\r)}, \qquad 
A_-=\frac{(\delm t_R+\cosh 2\r \delm t_L)}{(1+\cosh 2\r)}
\ee
Substituting into the action, 
and gauge fixing by setting $t_R=-t_L=t$, it is simple to see that we
get the black hole sigma model as mentioned.
We can instead reparametrise the gauge fields as $A_+=\del_+\phi_R$
and $A_-=\del_-\phi_L$ in terms of two noncompact scalars.
The reason these are noncompact has to do with the noncompactness of the 
symmetry being gauged. Upon shifting $t_{R,L}$ by
$\phi_{R,L}$ respectively, the Lagrangian becomes
\be
S=S_{WZNW}(t_L+\phi_L, \r, t_R+\phi_R) -\frac{k}{4\pi}\int d^2\sigma \sqrt{-h}\del_\alpha X\del^\alpha X
\ee
where $X=\ha (\phi_L-\phi_R)$. A quick way to see this is to start with the WZW model and then shift $t_{R,L}$.




Thus the classical stress tensor of this model is
\be
T_{++}=\frac{1}{k} \eta_{ab}J_+^a J_+^b - k(\del_+X)^2
\ee
and a similar expression for the $--$ components.
Upon quantizing the $1/k$ term becomes $1/(k-2)$ coming from the
standard WZNW renormalization. If we incorporate this first , and then
undo the field redefinitions, we obtain the sigma model metric to all
orders in $\a'$ \cite{Tseytlin}.

Although the above Lagrangian is the sum of two non-interacting theories,
there is a constraint relating the two. The constraint expressed in words
implies the vanishing of the total current of the $H$-subgroup that is being
gauged \cite{Karabali}.
This constraint is easily obtained. The equations of motion of the gauge
fields (which are really Lagrange multipliers enforcing the constraints) can
be written as
\be
J_-^{(2)}= k A_-(1+\cosh2\rho)\hspace{1cm}J_+^{(2)}=-kA_+(1+\cosh2\rho)
\ee
Shifting $t_{L,R}$ as before and parametrising $A_\pm=\del_\pm\phi_{L,R}$, the
above equations become
\be
J_\pm ^{(2)}=k\del_\pm X\label{constraint2}
\ee
which are the BRST constraints.

The level matching condition for the noncompact scalar $X$ forces the
conclusion that the left and right $J^{(2)}$ quantum numbers ought to
be equal.

\subsection{Generators}
We will choose the generators of \sl2 as in \cite{mog}. However, we
will label them so that $T^0$ is associated with the time direction of
the \sl2 geometry as
\be T^0=\frac{i}{2}\s^2 \quad T^2=\ha \s^3\,\,\,T^1=\ha\s^1\,\,\,\ee 
With this labelling, the generator $T^0$ is anti-hermitean.  However,
the black hole time is related to the $T^2$ direction and hence the
energy will still be real.  The commutation relations are
\be [T^2,T^1]=-T^0,\quad [T^2,T^0]=-T^1,\quad [T^1,T^0]=T^2. \ee 
In order to map these commutation relations to those in \cite{Mukunda} (note
that the generators in this latter work are hermitean), we make the
following identification 
\be T^2=J_1, \quad T^1=J_2, \quad T^0= iJ_3 \ee 

We will define the \sl2 currents as 
\be J^a _+=-k Tr(T^a \del_+ g g^{-1})\,\,\,\, J^a _-=k Tr(T^a g^{-1}\del_- g) \ee 
Given these conventions, the direction being gauged is $J^{(2)} _+ +
J^{(2)}_-$ and the black hole time direction is $J^{(2)}_+ - J^{(2)}_-$
The raising and lowering operators are $T^\pm=T^1 \pm T^0,$
respectively. But the commutation relations are a bit unusual
and given by $[T^2,T^\pm]=\pm i T^\pm$.

\subsection{Euclidean Black hole \label{EBH}}
In this section, we briefly illustrate how the winding strings of the
cigar geometry can be obtained by operation of spectral flow.  In this
case, we gauge the axial symmetry $g\rightarrow e^{i\sigma_2
  \Lambda}\,\,g e^{i\sigma_2 \Lambda}$ of the global \sl2 coordinates
$g= e^{i\sigma_2 (t+\q)}e^{\sigma_1\r}e^{i\sigma_2 (t-\q)}$.  The
coset geometry is obtained from gauge-fixed \sl2 matrices of the form
\[g=\cosh\rho+\sinh\rho\left(\barr\cos\theta&\sin\theta\\
-\sin\theta&\cos\theta\earr\right). \]
In this case, we perform the same ``spectral flow operation'' as in \sl2 \cite{mog}
\[ g\rightarrow
e^{i\omega_L\sigma_2 \sigma^+}\,g e^{-i\omega_R\sigma_2 \sigma^-}.\]
The axial gauge symmetry can be used to eliminate part of the spectral flow - but not the
whole of it. The gauge invariant content of the spectral flow is
\[ g\rightarrow
e^{i(m\tau+n\sigma)\sigma_2}\,g e^{-i(m\tau+n\sigma)\sigma_2}\]
under which $\theta\to\theta-m\tau+n\sigma$.  It is clear that the winding
strings come from this `spectrally flowed' sector.

\end{document}